\documentclass[11pt,notitlepage]{article}
\usepackage{amsmath,amsfonts,amssymb,amsthm,epsfig,epstopdf,url,array}
\usepackage{lscape}
\usepackage{geometry}
\usepackage[onehalfspacing]{setspace}
\usepackage{graphics}
\usepackage{graphicx}
\usepackage{subfig}
\usepackage{fancyhdr}
\usepackage{enumitem}
\usepackage{color}
\usepackage{caption}
\usepackage{titlesec}
\usepackage{floatrow}
\usepackage{tablefootnote}
\usepackage{bbm}
\theoremstyle{plain}

\newtheorem{prop}{Proposition}

\theoremstyle{definition}
\newtheorem{defn}{Definition}

\theoremstyle{remark}

\newenvironment{tablenotes}[1][0.5]{\begin{minipage}[t]{\linewidth}\footnotesize}{\end{minipage}}
\newenvironment{figurenotes}[1][0.5]{\begin{minipage}[t]{\linewidth}\footnotesize}{\end{minipage}}

\begin{document}
		
\pagestyle{fancy}
\chead{\small\sc Climate change heterogeneity}
\rhead{\small\thepage}
\renewcommand{\baselinestretch}{1}

\title{Climate change heterogeneity:\\ A new quantitative approach
	\date{July 10, 2022}
	\thanks{
	{\scriptsize The authors gratefully acknowledge the financial support from the Gobierno de Aragon and FEDER funds (grant, LMP71-18), the Spanish
		Ministerio de Ciencia y Tecnolog\'{\i}a, Agencia Espa\~{n}ola de Investigaci%
		\'{o}n (AEI) and European Regional Development Fund (ERDF, EU) under grants PID2019-104960GB-IOO,
		ECO2017-83255-C3-1-P (AEI/ERDF, EU) and ECO2016-81901-REDT, and Bank of Spain (ER grant program). We thank Rodrigo Gonzalez Laiz for excellent research assistance.}}}
\author{ Mar\'{\i}a Dolores Gadea Rivas
	\thanks{\scriptsize{
			Department of Applied Economics, University of Zaragoza. Gran V\'{\i}a, 4,
			50005 Zaragoza (Spain). Tel: +34 9767 61842,\ fax: +34 976 761840 and
			e-mail: lgadea@unizar.es}} \\
	{\small University of Zaragoza}
	\and Jes\'{u}s Gonzalo
	\thanks{\scriptsize{
			Department of Economics, University Carlos III, Madrid 126 28903 Getafe (Spain). Tel: +34 91 6249853,\ fax: +34 91 6249329 and e-mail:
			jesus.gonzalo@uc3m.es (corresponding author)}} \\
	{\small U. Carlos III de Madrid}}
\vspace*{-3cm}
{\let\newpage\relax\maketitle}

\begin{abstract}
{\small \noindent \\}
Climate change is a non-uniform phenomenon. This paper proposes a new quantitative methodology to characterize, measure and test the existence of climate change heterogeneity. It consists of three steps. First, we introduce a new testable warming typology based on the evolution of the trend of the whole temperature distribution and not only on the average. Second, we define the concepts of warming acceleration and warming amplification in a testable format. And third, we introduce the new testable concept of warming dominance to determine whether region \textit{A} is suffering a worse warming process than region \textit{B}. Applying this three-step methodology, we find that Spain and the Globe experience a clear distributional warming process (beyond the standard average) but of different types. In both cases, this process is accelerating over time and asymmetrically amplified. Overall, warming in Spain dominates the Globe in all the quantiles except the lower tail of the global temperature distribution that corresponds to the Artic region. Our climate change heterogeneity results open the door to the need for a non-uniform causal-effect climate analysis that goes beyond the standard causality in mean as well as for a more efficient design of the mitigation-adaptation policies. In particular, the heterogeneity we find suggests that these policies should contain a common global component and a clear local-regional element. Future climate agreements should take the whole temperature distribution into account.
\\

{\small \noindent \textit{JEL classification}: C31, C32, Q54}

{\small \noindent \textit{Keywords}: Climate change; Climate heterogeneity; Global-Local Warming;  Functional stochastic processes; Distributional characteristics; Trends; Quantiles;  Temperature distributions.}

\end{abstract}

\newpage
\section{Introduction\label{sec-introduction}}
﻿
All the assessment reports (AR) published by the Intergovernmental Panel of Climate Change (IPCC) show that there is overwhelming scientific evidence of the existence of global warming (GW). It is also well known that climate change (CC) is a non-uniform phenomenon. What is not so clear is the degree of heterogeneity across all the regions in our planet. In fact, an important part of the Sixth Assessment Report (AR6) published by the IPCC in 2021-2022 is dedicated to this issue: climate (warming) heterogeneity. This is reflected in the chapters studying regional climate change. Our paper introduces a new quantitative methodology that builds on that described in Gadea and Gonzalo 2020 (GG2020) to characterize, measure and test the existence of such climate change heterogeneity (CCH). This is done in three steps. First, we introduce a warming typology (\textit{W1}, \textit{W2} and \textit{W3}) based on the trending behavior of the quantiles of the temperature distribution of a given geographical location. Second, we define in a testable format the concepts of warming acceleration and warming amplification. These concepts help to characterize (more ordinally than cardinally) the warming process of different regions. And third, we propose the new concept of warming dominance (WD) to establish when region \textit{A} suffers a worse warming process than region \textit{B}.

We have chosen Spain as a benchmark geographical location because, as the AR6 report states “. . . Spain is fully included in the Mediterranean (MED) Reference Region, but is one of the most climatically diverse countries in the world. . . ”. This fact opens up the possibility of studying warming heterogeneity (WH) from Spain to the Globe (outer heterogeneity, OWH) and also from Spain to some of its regions represented by Madrid and Barcelona (inner heterogeneity, IWH).

The three steps rely on the results reported in GG2020, where the different distributional characteristics (moments, quantiles, inter quantile range, etc.) of the temperature distribution of a given geographical location are converted into time series objects. By doing this, we can easily implement and test all the concepts involved in the  three steps.

A summary of the results is as follows. Spain and the Globe present a clear warming process; but it evolves differently. Spain goes from a warming process where lower and upper temperatures share the same trend behavior (\textit{IQR} is maintained constant over time, warming type \textit{W1}) to one characterized by a larger increase in the upper temperatures (\textit{IQR} increases over time, warming type \textit{W3}). In contrast, the  Globe as a whole maintains a stable warming type process characterized by lower temperatures that increase more than the upper ones (\textit{IQR} decreases in time).\footnote{Similar results for Central England are found in GG2020 and for the US in Diebold and Rudebush, 2022.} In our typology, this constitutes a case of warming type \textit{W2}. Climate heterogeneity can go further. For instance, within Spain we find that Madrid is of type \textit{W3} while the warming process of Barcelona is of type \textit{W1}. This is in concordance with the Madrid climate being considered a Continental Mediterranean climate while Barcelona is more a pure Mediterranean one.

The proposed warming typology (\textit{W1}, \textit{W2} and \textit{W3}), although dynamic, is more ordinal than cardinal. In this paper, the strength of a warming process is captured in the second step by analyzing its acceleration and its amplification with respect to a central tendency measure of the temperature distribution. Acceleration and amplification contribute to the analysis of warming heterogeneity. The acceleration in the Globe is present in all the quantiles above \textit{q30} while in Spain it already becomes significant above the 10$^{th}$ quantile. We find an asymmetric behavior of warming amplification; in Spain (in comparison with the Globe mean temperature) this is present in the upper temperatures (above the 80$^{th}$ and 90$^{th}$ quantiles) while in the Globe the opposite occurs (below the 20$^{th}$ and 30$^{th}$ quantiles). Within Spain, Madrid and Barcelona also behave differently in terms of acceleration and amplification. Overall, warming in Spain dominates that of the Globe in all the quantiles except for the lower quantile \textit{q05}, and between Madrid and Barcelona there is a partial WD. Madrid WD Barcelona in the upper part of the distribution and Barcelona WD Madrid in the lower one.

The existence of a clear heterogeneous warming process opens the door to the need of a new non-uniform causal (effect) research. One that goes beyond the standard causality in mean analysis (see Tol, 2021). CCH also suggests that in order for the
mitigation-adaptation policies to be as efficient as possible they should be designed following a type of common factor structure: a common global component plus an idiosyncratic local element. This goes in the line with the results found in  Brock and  Xepapadeas (2017), D’Autume  et al. (2016) and Peng et al. (2021). Future climate agreements should clearly have this CCH into account. An important by-product of our warming heterogeneity results is the increase that this heterogeneity can generate in the public awareness of the GW process. A possible explanation for that can be found in the behavioral economics work by Malmendier (2021), in the results of the European Social Survey analyzed in Nowakowski and Oswald (2020) or in the psychology survey by Maiella et al. (2020).

The rest of the paper is organized as follows. Section 2 describes our basic climate econometrics methodology. Section 3 presents a brief description of the temperature data from Spain and the Globe. Section 4 addresses the application of our quantitative methodology in the cross-sectional version (temperatures measured monthly by stations in an annual interval) to Spain and (versus) the Globe. It also reports the results of applying the methodology using a purely temporal dimension (local daily temperature on an annual basis) for two representative stations in Spain (Madrid and Barcelona, empirical details in the Appendix). Section 5 offers a comparison and interpretation of the results. Finally, Section 6 concludes the paper.

\section{Climate Econometrics Methodology\label{sec-method}}
In this section, we briefly summarize the novel econometric methodology introduced in GG2020 to analyze Global and Local Warming processes.
Following GG2020, Warming is defined as an increasing trend in certain characteristics of the temperature distribution. More precisely:
\begin{defn}  \label{def1} \textit{(\underline{Warming})}:
\textit{	Warming is defined as the existence of an increasing trend in some of the characteristics measuring the central tendency or position (quantiles) of the temperature distribution.}  	
\end{defn}

An example is a deterministic trend with a polynomial function for certain values of the $\beta$ parameters $C_{t}=\beta _{0}+\beta _{1}t+\beta _{2}t^{2}+...+\beta _{k}t^{k}$. \\

In GG2020 temperature is viewed as a functional stochastic process  $X=(X_{t}(\omega), t \in T)$, where $T$ is an interval in $\mathbb{R}$, defined in a probability space $(\Omega, \Im, P)$. A convenient example of an infinite-dimensional discrete-time process consists of associating $\xi=(\xi_n, n \in \mathbb{R}_{+})$ with a sequence of random variables whose values are in an appropriate function space. This may be obtained by setting
\begin{equation}
	X_{t}(n)=\xi_{tN+n}, \text{   } 0\leq n \leq N, \text{   } t=0,1,2, ..., T \label{example}
\end{equation}
so $X=(X_{t}, t=0,1,2,...,T)$. If the sample paths of $\xi$ are continuous, then we have a sequence $X_{0}, X_{1}, ....$ of random variables in the space $C[0, N]$. The choice of the period or segment $t$ will depend on the situation in hand. In our case, $t$ will be the period of a year, and $N$ represents cross-sectional units or higher-frequency time series.

We may be interested in modeling the whole sequence of $\mathbf{G}$ functions, for instance the sequence of state densities ($f_{1}(\omega), f_{2}(\omega), ..., f_{T}(\omega) $ ) as in Chang et al. (2015, 2016) or only certain characteristics ($C_{t}(w)$) of these $\mathbf{G}$ functions, for instance, the state mean, the state variance, the state quantile, etc. These characteristics can be considered time series objects and, therefore, all the econometric tools already developed in the time series literature can be applied to $C_{t}(w)$. With this characteristic approach we go from $\Omega$ to $\mathbb{R}^{T}$, as in a standard stochastic process, passing through a $\mathbf{G}$ functional space:
\begin{center}
	$\underset{(w)}{\Omega} \xrightarrow{X} \underset{X_{t}(w)}{\mathbf{G}} \xrightarrow{C} \underset{C_{t}(w)}{\mathbb{R}}$ \\
\end{center}

Going back to the convenient example and abusing notation, the stochastic structure can be summarized in the following array:

\begin{equation}
	\scalebox{0.8}{
		\begin{tabular}{|c|cc|c|c|}
			\hline
			${\small X}_{{\small 10}}{\small (w)=}$ $\xi _{0}(w)$ & \multicolumn{1}{|c|}{%
				${\small X}_{{\small 11}}{\small (w)=}$ $\xi _{1}(w)$} & $.$ $.$ $.$ &
			\multicolumn{1}{|c|}{${\small X}_{{\small 1N}}{\small (w)=}$ $\xi _{N}(w)$}
			& ${\small C}_{{\small 1}}{\small (w)}$ \\ \hline
			${\small X}_{{\small 20}}{\small (w)=}$ $\xi _{N+1}(w)$ &
			\multicolumn{1}{|c|}{${\small X}_{{\small 21}}{\small (w)=}$ $\xi _{N+2}(w)$}
			& $.$ $.$ $.$ & \multicolumn{1}{|c|}{${\small X}_{{\small 2N}}{\small (w)=}$
				$\xi _{2N}(w)$} & ${\small C}_{{\small 2}}{\small (w)}$ \\ \hline
			$%
			\begin{array}{c}
				. \\
				. \\
				.%
			\end{array}%
			$ & \multicolumn{1}{|c|}{$%
				\begin{array}{c}
					. \\
					. \\
					.%
				\end{array}%
				$} & $%
			\begin{array}{c}
				.\text{ }.\text{ }. \\
				.\text{ }.\text{ }. \\
				.\text{ }.\text{ }.%
			\end{array}%
			$ & \multicolumn{1}{|c|}{$%
				\begin{array}{c}
					. \\
					. \\
					.%
				\end{array}%
				$} & $%
			\begin{array}{c}
				. \\
				. \\
				.%
			\end{array}%
			$ \\ \hline
			${\small X}_{{\small T0}}{\small (w)=}$ $\xi _{(T-1)N+1}(w)$ &
			\multicolumn{1}{|c|}{${\small X}_{{\small T1}}{\small (w)=}$ $\xi
				_{(T-1)N+2}(w)$} & $.$ $.$ $.$ & \multicolumn{1}{|c|}{${\small X}_{{\small TN%
				}}{\small (w)=}$ $\xi _{TN}(w)$} & ${\small C}_{{\small T}}{\small (w)}$ \\
			\hline
		\end{tabular}
	}
	\label{eq-scheme}
\end{equation}

The objective of this section is to provide a simple test to detect the existence of a general unknown trend component in a given characteristic $C_t$ of the temperature process $X_t$. To do this, we need to convert Definition \ref{def1} into a more practical definition.

\begin{defn} \label{def2} \textit{(\underline{Trend test})}:  \textit{Let $h(t)$ be an increasing function of $t$. A characteristic $C_{t}$ of a functional stochastic process $X_{t}$ contains a trend  if $\beta \neq 0$ in the regression}
	\begin{equation}
		C_{t}=\alpha +\beta h(t)+u_{t}, \text{   } t=1,...,T. \label{tbeta}
	\end{equation}
\end{defn}

The main problem of this definition is that the trend component in $C_t$ as well as the function $h(t)$ are unknown. Therefore this definition can not be easily implemented. If we assume that $C_t$ does not have a trend component (it is $I(0)$)\footnote{Our definition of an I(0) process follows Johansen (1995). A stochastic process $Y_{t}$ that satisfies $Y_{t}-E(Y_{t})$ $%
	=\sum \limits_{i=1}^{\infty }\Psi_{i}\varepsilon _{t-i}$ is called I(0) if $%
	\sum \limits_{i=1}^{\infty }\Psi$ $_{i}z^{i}$ converges for $\left \vert
	z\right \vert <1+\delta$, for some $\delta>0$ and $\sum \limits_{i=1}^{\infty }\Psi$ $_{i}\neq 0,$ where the
	condition $\varepsilon_{t}\thicksim $ iid(0,$\sigma ^{2})$ with $\sigma ^{2}>0$ is understood.} and $h(t)$ is linear, then we have the following well known result.

\begin{prop}\label{prop1}
	Let $C_{t}=I(0)$. In the regression
	\begin{equation}
		C_{t}=\alpha +\beta t + u_{t}
		\label{eq-reg}
	\end{equation}
	the OLS estimator
	\begin{equation}
		\widehat{\beta}=\frac{\sum \limits_{t=1}^{T}(C_{t}-\overline{C})(t-\overline{t})}{\sum \limits_{t=1}^{T}(t-\overline{t})^{2}}
	\end{equation}
	satisfies
	\begin{equation}
		T^{3/2}\widehat{\beta }=O_{p}(1)
	\end{equation}
	and asymptotically ($T \rightarrow \infty$)
	\begin{equation*}
		t_{\beta =0} \text{  is  }  N(0,1).
	\end{equation*}
\end{prop}

In order to analyze the behavior of the t-statistic $t_{\beta }=0,$ for a general trend component in $C_t$, it is very convenient to use the concept of \textit{Summability} (Berenguer-Rico and Gonzalo, 2014)

\begin{defn} \label{def3} \textit{(\underline{Order of Summability})}: \textit{ A trend $h(t)$ is said to be summable of order ``$\delta$'' $(S(\delta ))$ if there exists a slowly varying function $L(T)$,\footnote{A positive Lebesgue measurable function, L, on $(0,\infty)$ is slowly varying (in Karamata's sense) at $\infty$ if
		\begin{equation}
			\frac{L(\lambda n)}{L(n)}\rightarrow 1\text{ }(n\rightarrow \infty )\text{ }%
			\forall \lambda >0.
		\end{equation}
		(See Embrechts et al., 1999, p. 564).} such that}
	
	\begin{equation}
		S_{T}=\frac{1}{T^{1+\delta }}L(T)\sum_{t=1}^{T}h(t)  \label{eq_sum}
	\end{equation}
	\textit{is $O(1)$, but not $o(1)$.}
\end{defn}

\begin{prop}\label{prop2}
	Let $C_{t}=h(t)+I(0)$ such that $h(t)$ is $ S(\delta )$ with $\delta \geq 0$, and such that the function $g(t)=h(t)t $ is $ S(\delta +1)$.
	In the regression
	\begin{equation}
		C_{t}=\alpha +\beta t + u_{t} \label{tbeta2}
	\end{equation}
	the OLS $\widehat{\beta}$ estimator satisfies
	\begin{equation}
		T^{(1-\delta )}\widehat{\beta }=O_{p}(1).
	\end{equation}
	
	Assuming that the function $h(t)^{2}$ is $ S(1+2 \delta-\gamma)$ with $0\leq \gamma \leq1+\delta $, then
	
	\begin{equation}
		t_{\beta =0} = \left \{
		\begin{array}{l}
			O_{p}(T^{\gamma/2})$ for $0\leq \gamma \leq1 \\
			O_{p}(T^{1/2})$ for $1\leq \gamma \leq1+\delta
		\end{array}  \right.
	\end{equation}
\end{prop}

Examples of how this proposition applies for different particular Data Generating Processes (DGP) can be found in GG.\\

A question of great empirical importance is how our trend test ($TT$) of Proposition \ref{prop2} behaves when $C_t=I(1)$  (accumulation of an I(0) process). Following Durlauf and Phillips (1988), $T^{1/2}\widehat{\beta}=O_{p}(1)$; however, $t_{\beta =0}$ diverges as $ T {\rightarrow } \infty$. Therefore, our $TT$ can detect the stochastic trend generated by an I(1) process. In fact, our test will detect trends generated by any of the three standard persistent processes considered in the literature (see Muller and Watson, 2008): (i) fractional or long-memory models; (ii) near-unit-root AR models; and (iii) local-level models. Let

\begin{equation}
	C_{t}=\mu+z_{t},\text{   } t=1,...,T.  \label{eq-sto_trend}
\end{equation}

In the first model, $z_{t}$ is a fractional process with $1/2<d<3/2$. In the second model, $z_{t}$ follows an AR, with its largest root close to unity, $\rho _{T}=1-c/T$. In the third model, $z_{t}$ is decomposed into an I(1) and an I(0) component. Its simplest format is $z_{t}$ = $\upsilon _{t}$ + $\epsilon _{t}$ with $\upsilon _{t}$ = $\upsilon _{t-1}$ +$\eta _{t}$, where $\epsilon _{t}$ is $ID(0,q\ast \sigma ^{2}$), $\eta _{t}$ is $ID(0,\sigma ^{2})$, $\sigma^{2} >0$ and both disturbances are serially and mutually independent. Note that the pure unit-root process is nested in all three models: $d=1$, $c=0$, and $q=0$.

The long-run properties implied by each of these models can be characterized using the stochastic properties of the partial sum process for $z_{t}$. The standard assumptions considered in the macroeconomics or finance literature assume the existence of a ``$\delta$,'' such that $T^{-1/2+\delta }\sum_{t=1}^{T}z_{t}\longrightarrow \sigma $ $H(.)$, where ``$\delta$'' is a model-specific constant and $H$ is a model-specific zero-mean Gaussian process with a given covariance kernel $k(r,s).$ Then, it is clear that the process $C_{t}=\mu+z_{t}$ is summable (see Berenguer-Rico and Gonzalo, 2014). This is the main reason why Proposition \ref{prop3} holds for these three persistent processes.

\begin{prop}\label{prop3}
	Let $C_{t}=\mu+z_{t},t=1,...,T$, with $z_{t}$ any of the following three processes: (i) a fractional or long-memory model, with $1/2<d<3/2$; (ii) a near-unit-root AR model; or (iii) a local-level model. Furthermore, $T^{-1/2+\delta }\sum_{t=1}^{T}z_{t}\longrightarrow \sigma $ $H(.)$,
	where ``$\delta$'' is a model-specific constant and $H$ is a model-specific zero-mean Gaussian process with a given covariance kernel $k(r,s).$
	Then, in the LS regression
	\begin{equation*}
		C_{t}=\alpha+\beta t+u_{t},
	\end{equation*}
	the t-statistic diverges,
	\begin{equation*}
		t_{\beta =0}=O_{p}(T^{1/2}).
	\end{equation*}
	
\end{prop}

After the development of the theoretical core, we are in a position to design tools to approach the empirical strategy. The following subsection describes each of them.

\subsection{Empirical tools: definitions and tests}
From Propositions \ref{prop2} and \ref{prop3}, Definition \ref{def2} can be simplified into the following testable and practical definition.
\begin{defn} \label{def4} \textit{(\underline{Practical definition 2})}: \textit{ A characteristic $C_{t}$ of a functional stochastic process $X_{t}$ contains a trend if in the LS regression,}
	\begin{equation}
		C_{t}=\alpha +\beta t+u_{t}, \text{   } t=1,...,T,  \label{tbeta3}
	\end{equation}
	\textit{$\beta=0$ is rejected.}
\end{defn}

Several remarks are relevant with respect to this definition: (i) regression (\ref{tbeta3}) has to be understood as the linear LS approximation of an unknown trend function $h(t)$ (see White, 1980); (ii) the parameter $\beta$ is the plim of $\widehat{\beta}_{ols}$; (iii) if the regression (\ref{tbeta3}) is the true data-generating process, with $u_t\sim I(0)$, then the OLS $\widehat{\beta }$ estimator is asymptotically equivalent to the GLS estimator (see Grenander and Rosenblatt, 1957);  (iv) in practice, in order to test $\beta=0$, it is recommended to use a robust HAC version of $t_{\beta =0}$ (see Busetti and Harvey, 2008); and (v) this test only detects the existence of a trend but not the type of trend.

For all these reasons, in the empirical applications we implement Definition \ref{def4} by estimating regression (\ref{tbeta3}) using OLS and constructing a HAC version of $t_{\beta =0}$ (Newey and West, 1987).

These linear trends can be common across characteristics indicating similar patters in the time evolution of these characteristics.

\begin{defn} \label{def5} \textit{(\underline{Co-trending})}:  \textit{A set of $m $ distributional characteristics ($C_{1t}$,$C_{2t}$,...,$C_{mt}$) do linearly co-trend if in the multivariate regression \\}

	\begin{equation}
		\begin{pmatrix}
			C_{1t} \\
			... \\
			C_{mt}%
		\end{pmatrix}%
		=%
		\begin{pmatrix}
			\alpha _{1} \\
			... \\
			\alpha _{m}%
		\end{pmatrix}%
		+%
		\begin{pmatrix}
			\beta _{1} \\
			... \\
			\beta _{m}%
		\end{pmatrix}%
		t+%
		\begin{pmatrix}
			u_{1t} \\
			... \\
			u_{mt}%
		\end{pmatrix}%
		\label{cotrend}
	\end{equation}
	
\textit{	all the slopes are equal, $\beta _{1}=\beta _{2}=...=\beta _{m}.$} \footnote{This definition is slightly different from the one in Carrion-i-Silvestre and  Kim (2019).}
\end{defn}

This co-trending hypothesis can be tested by a standard Wald test.

When $m=2$ an alternative linear co-trending test can be obtained from
the regression
\begin{equation*}
	C_{it}-C_{jt}=\alpha +\beta t+u_{t}
\end{equation*}				
$i\neq j$ $i,j=1,...,m$ by testing the null hypothesis of $\beta =0$ vs $\beta \neq 0$ using
a simple $t_{\beta =0}$ test.

Climate classification is a tool used to recognize, clarify and simplify the existent climate heterogeneity in the Globe. It also helps us to better understand the Globe’s climate and therefore to design more efficient global warming mitigation policies. The prevalent climate typology is that proposed by K\"oppen (1900) and later on modified in K\"oppen and Geiger (1930). It is an empirical classification that divides the climate into five major types, which are represented by the capital letters A (tropical zone), B (dry zone), C (temperate zone), D (continental zone), and E (polar zone). Each of these climate types except for B is defined by temperature criteria. More recent classifications can been found in the AR6 of the IPCC (2021, 2022) but all of them share the spirit of the original one of K\"oppen (1900).

The climate classification we propose in this section is also based on temperature data and it has three simple distinctive characteristics:
\begin{itemize}
	\item It considers the whole temperature distribution and not only the average
	\item It has a dynamic nature: it is based on the evolution of the trend of the temperature quantiles (lower and upper).
	\item It can be easily tested
\end{itemize}

\begin{defn}  \label{def6} \textit{(\underline{Warming Typology})}:
\textit{We define four types of warming processes:}
	
	\begin{itemize}
		\item \textbf{W0}: \textit{There is no trend in any of the quantiles (No warming).}
		\item \textbf{W1}: \textit{All the location distributional characteristics have the same positive trend (dispersion does not contain a trend)}
		\item \textbf{W2}: \textit{The Lower quantiles have a larger positive trend than the Upper quantiles (dispersion has a negative trend)}
		\item \textbf{W3}: \textit{The Upper quantiles have a larger positive trend than the Lower quantiles (dispersion has a positive trend).}
	\end{itemize}	
\end{defn}

Climate is understood, unlike weather, as a medium and long-term phenomenon and, therefore, it is crucial to take trends into account. Notice that this typology can be used to describe macroclimate as well as microclimate locations.

Most of the literature on Global or Local warming only considers the trend behavior of the central part of the distribution (mean or median). By doing this, we are losing very useful information that can be used to describe the whole warming process. This information is considered in the other elements of the typology \textit{W1}, \textit{W2} and \textit{W3}. This typology does not say anything about the intensity of the warming process and its dynamic. Part of this intensity is captured in the following definitions of warming acceleration and warming amplification.

\begin{defn}  \label{def7} \textit{(\underline{Warming Acceleration})}:
	\textit{We say that there is warming acceleration in a distributional temperature characteristic $C_{t}$ between the time periods $t_1=(1,..., s)$ and $t_2=(s+1,..., T)$ if in the following two regressions:
}
\begin{equation}
	C_{t}=\alpha_{1} +\beta_{1} t+u_{t}, \text{   } t=1, ...,s ,..., T,
\end{equation}
\begin{equation}
	C_{t}=\alpha_{2} +\beta_{2} t+u_{t}, \text{   } t=s+1, ..., T, \label{acc}
\end{equation}
\textit{the second trend slope is larger than the first one: $\beta_{2} > \beta_{1}$.}\\
\end{defn}

In practice, we implement this definition by testing in the previous system the null hypothesis $\beta_{2}=\beta_{1}$ against the alternative $\beta_{2}>\beta_{1}$ An alternative warming acceleration test can be formed by testing for a structural break at $t=s$. Nevertheless, we prefer the approach of Definition \ref{def7} because it matches closely the existent narrative on warming acceleration in the climate literature.

\begin{defn}  \label{def8} \textit{(\underline{Warming Amplification with respect to the mean})}:
\textit{	We say that there is a warming amplification in distributional characteristic $C_{t}$ with respect the $mean$ if in the following regression:}	
\begin{equation}
		C_{t}=\beta _{0}+\beta _{1} mean_{t}+\epsilon_{t} \label{ampl}
	\end{equation}
\textit{the mean slope is greater than one: $\beta_{1} >1$.	}
\end{defn}	
When the mean, $mean_{t}$, and $C_{t}$ come from the same distribution, we name this ``inner'' warming amplification. Otherwise, the mean may come from an external environment and, in that case, we call it ``outer'' warming amplification.

Both concepts, acceleration and amplification, introduce a quantitative dimension to the ordinarily defined classification. For example, the acceleration, which has a dynamic character, allows us to observe the transition from one type of climate to another. Amplification, on the other hand, makes it possible to compare the magnitude of the trends that define each type of climate. It should be noted that, although static in nature, it can be computed recursively at different points in time.

In the previous definitions, we classify the warming process of different regions which is crucial in the design of local mitigation and adaptation policies. But we, also, need to compare the different climate change processes of two regions in order to characterize climate heterogeneity independently of the type of warming they are experimenting. For this purpose, we propose the following definition that shares the spirit of the stochastic dominance concept used in the economic-finance literature.

	\begin{defn}  \label{def9} \textit{(\underline{Warming Dominance (WD)}}:
\textit{We say that the temperature distributions of \textbf{Region $A$} warming dominates (\textbf{$WD$}) the temperature distributions of \textbf{Region $B$} if in the following regression
}	
	\begin{equation}
		q_{\tau t}(A)- q_{\tau t}(B)=\alpha_{\tau} +\beta_{\tau} t +u_{\tau t} \label{wd},
	\end{equation}
	
\textit{$\beta_{\tau}\geq 0$ for all $0<\tau<1$ and there is at least one value $\tau^{*}$ for which a strict inequality holds.}
\end{defn}	
It is also possible to have only \emph{partial} (\textbf{$WD$}). For instance, in the lower or upper quantiles.

\section{The data\label{sec-data}}
\subsection{Spain}
The measurement of meteorological information in Spain started in the eighteenth century. However, it was not until the mid-nineteenth century  that reliable and regular data became available. In Spain, there are four main sources of meteorological information: the Resumen Anual, Bolet\'{\i}n Diario, Bolet\'{\i}n Mensual de Climatolog\'{\i}a and Calendario Meteorol\'ogico. These were first published in 1866, 1893, 1940 and 1943, respectively. A detailed explanation of the different sources can be found in Carreras and Tafunell (2006).

Currently, AEMET (Agencia Estatal de Meterolog\'{\i}a) is the agency responsible for storing, managing and providing meteorological data to the public. Some of the historical publications, such as the Bolet\'{\i}n Diario and Calendario Meteorol\'ogico can be found in digital format in their respective archives for whose use it is necessary to use some kind of Optical Character Recognition (OCR) software.\footnote{$http://www.aemet.es/es/conocermas/recursos_en_linea/calendarios?n=todos$ and $https://repositorio.aemet.es/handle/20.500.11765/6290$.}

In 2015, AEMET developed AEMET OpenData, an Application Programming Interface (API REST) that allows the dissemination and reuse of Spanish meteorological and climatological information. To use it, the user needs to obtain an API key to allow access to the application. Then, either through the GUI or through a programming language such as Java or Python, the user can request data. More information about the use of the API can be found on their webpage.\footnote{$https://opendata.aemet.es/centrodedescargas/inicio$. The use of AEMET data is regulated in the following resolution $https://www.boe.es/boe/dias/2016/01/05/pdfs/BOE-A-2016-111.pdf$.}

In this paper, we are concerned with Spanish daily station data, specifically temperature data. Each station records the minimum, maximum and average temperature as well as the amount of precipitation, measured as liters per square meter. The data period ranges from 1920 to 2019. However, in 1920 there were only 13 provinces (out of 52) who had stations available. It was not until 1965 that all the 52 provinces had at least one working station. Moreover, it is important to keep in mind that the number of stations has increased substantially from only 14 stations in 1920 to more than 250 in 2019.
With this information in mind, we select the longest span of time that guarantees a wide sample of stations so that all the geographical areas of peninsular Spain are represented. For this reason, we decided to work with station data from 1950 to 2019. There are 30 stations whose geographical distribution is displayed in the map in Figure \ref{fig-data}. The original daily data are converted into monthly data, so that we finally work with a total of 30x12 station-month units corresponding to peninsular Spain and, consequently, we have 360 observations each year with which to construct the annual distributional characteristics.

\subsection{The Globe}
In the case of the Globe, we use the database of the Climate Research Unit (CRU) that offers monthly and yearly data of land and sea temperatures in both hemispheres from 1850 to the present, collected from different stations around the world.\footnote{We use CRUTEM version 5.0.1.0, which can be downloaded from (https://crudata.uea.ac.uk/cru/data/temperature/). A recent revision of the methodology can be found in Jones et al. (2012).} Each station temperature is converted to an anomaly, taking 1961-1990 as the base period, and each grid-box value, on a five-degree grid, is the mean of all the station anomalies within that grid box. This database (in particular, the annual temperature of the Northern Hemisphere) has become one of the most widely used to illustrate GW from records of thermometer readings. These records form the blade of the well-known ``hockey stick'' graph, frequently used by academics and other institutions, such as, the IPCC. In this paper, we prefer to base our analysis on raw station data, as in GG2020.

\begin{figure}[h!]
	\begin{center}
		\caption{Geographical distribution of stations}
		\label{fig-data}
		\subfloat[{\small Spain. Selected stations, AEMET data 1950-2019}]{
			\includegraphics[scale=0.9]{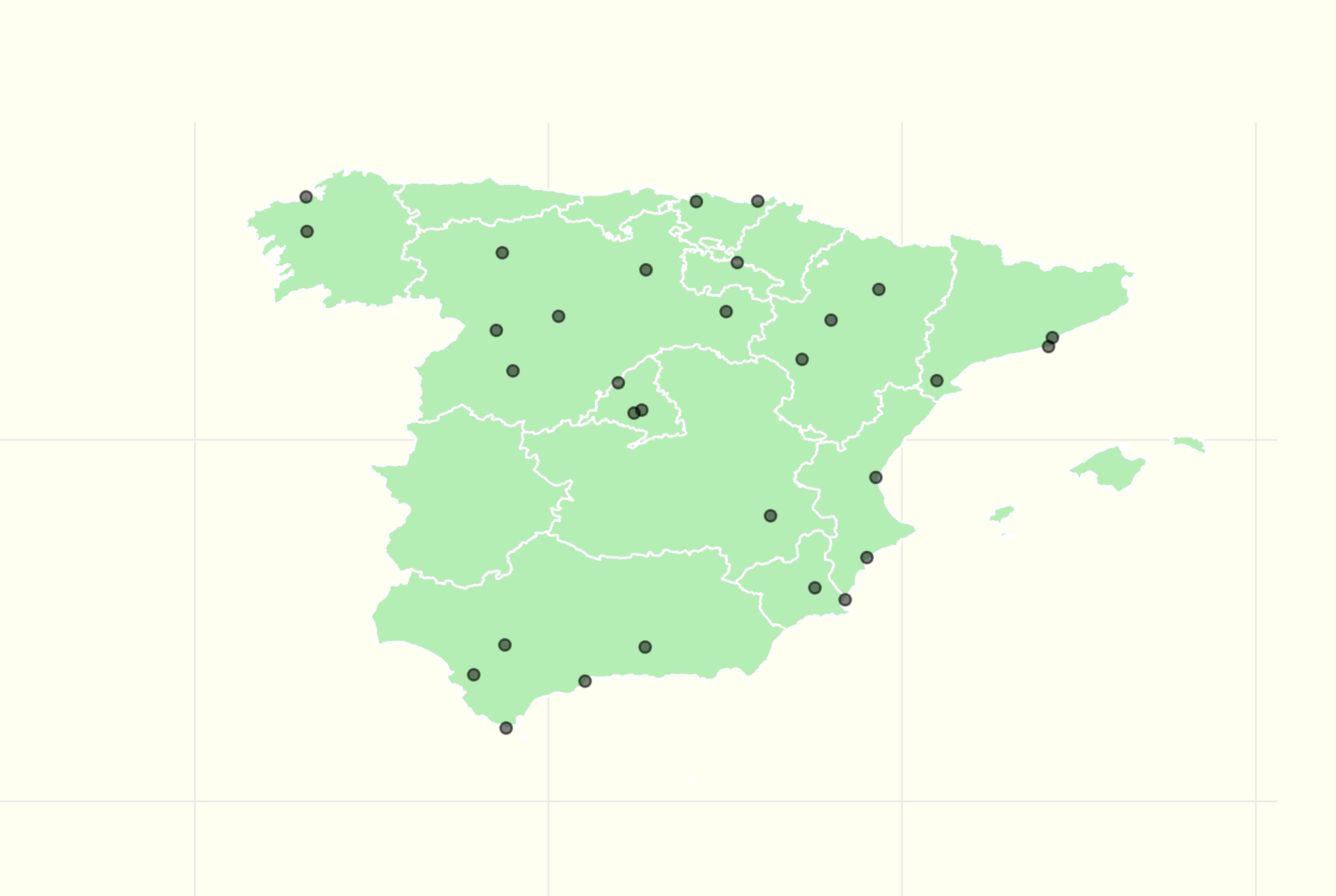}}\\

		\subfloat[{\small The Globe. Selected stations, CRU data 1950-2019}]{
			\includegraphics[scale=0.7]{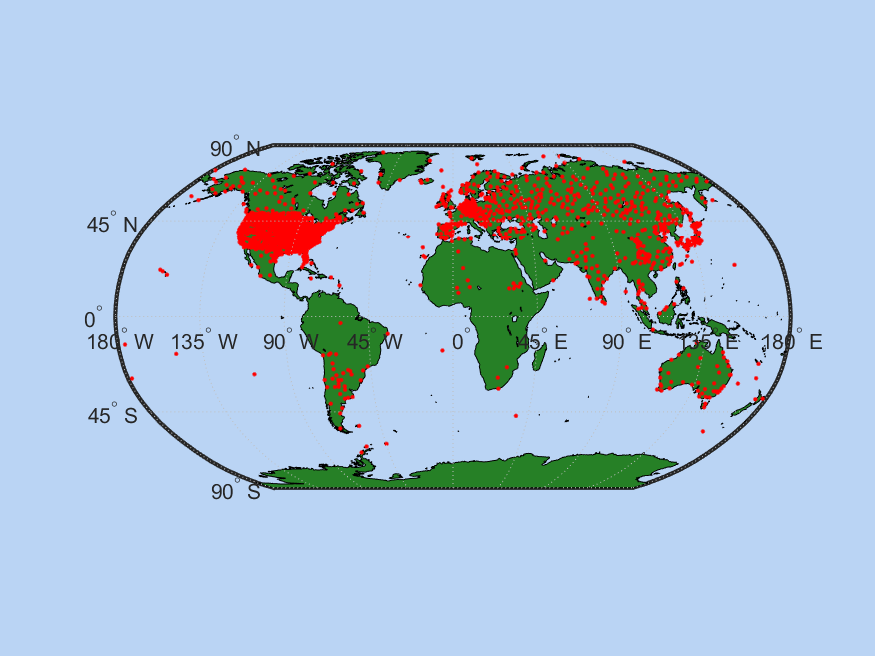}}
	\end{center}
\end{figure}

The database provides data from 1850 to nowadays, although due to the high variability at the beginning of the period it is customary in the literature to begin in 1880. In this work, we have selected the stations that are permanently present in the period 1950-2019 according to the concept of the station-month unit. In this way, the results are comparable with those obtained for Spain. Although there are 10,633 stations on record, the effective number fluctuates each year and there are only 2,192 stations with data for all the years in the sample period, which yields 19,284 station-month units each year (see this geographical distribution in the map in Figure \ref{fig-data}).\footnote{In the CRU data there are 115 Spanish stations. However, after removing stations not present for the whole 1880 to 2019 period, only Madrid-Retiro, Valladolid and Soria remain. Since 1950, applying the same criteria, only 30 remain.} In summary, we analyze raw global data (stations instead of grids) for the period 1950 to 2019, compute station-month units that remain all the time and with these build the annual distributional characteristics.

\section{Empirical strategy\label{sec-emp}}
In this section we apply our three-step quantitative methodology to show the existent climate heterogeneity between Spain and the Globe as well as within Spain, between Madrid and Barcelona. Because all our definitions are written in a testing format, it is straightforward to empirically apply them. First, we test for the existence of warming by testing the existence of a trend in a given distributional characteristic. How common are the trends of the different characteristics (revealed by a co-trending test) determine the warming typology. Second, the strength of the warming process is tested by testing the hypothesis of warming acceleration and warming amplification. And third, independently of the warming typology, we determine how the warming process of Spain compares with that of the Globe as a whole (we do the same for Madrid and Barcelona). This is done by testing for warming dominance.

The results are presented according to the following steps: first, we apply our trend test (see Definition \ref{def4}) to determine the existence of local or global warming and test for any possible warming acceleration; second, we test different co-trending hypotheses to determine the type of warming of each area; thirdly, we test the warming amplification hypothesis for different quantiles with respect to the mean (of Spain as well as of the Globe): $H_{0}: \beta_{1}=1$ versus $H_{a}: \beta_{1}>1$ in (\ref{ampl}); and finally, we compare the \textit{CC} of different regions, for Spain and the Globe, and within Spain, between Madrid and Barcelona, with our warming dominance test (see \ref{wd}).\footnote{Before testing for the presence of trends in the distributional characteristics of the data, we test for the existence of unit roots. To do so, we use the well-known Augmented Dickey-Fuller test (ADF; Dickey and Fuller, 1979), where the number of lags is selected in accordance with the SBIC criterion. The results, available from the authors on request, show that the null hypothesis of a unit root is rejected for all the characteristics considered.}
	
\subsection{Local warming: Spain \label{sec-cross-Spain}}
\begin{figure}[h!]
	\begin{center}
		\caption{Spain annual temperature density calculated with monthly data across stations} \label{fig-density-Spain}
		\includegraphics[scale=0.5]{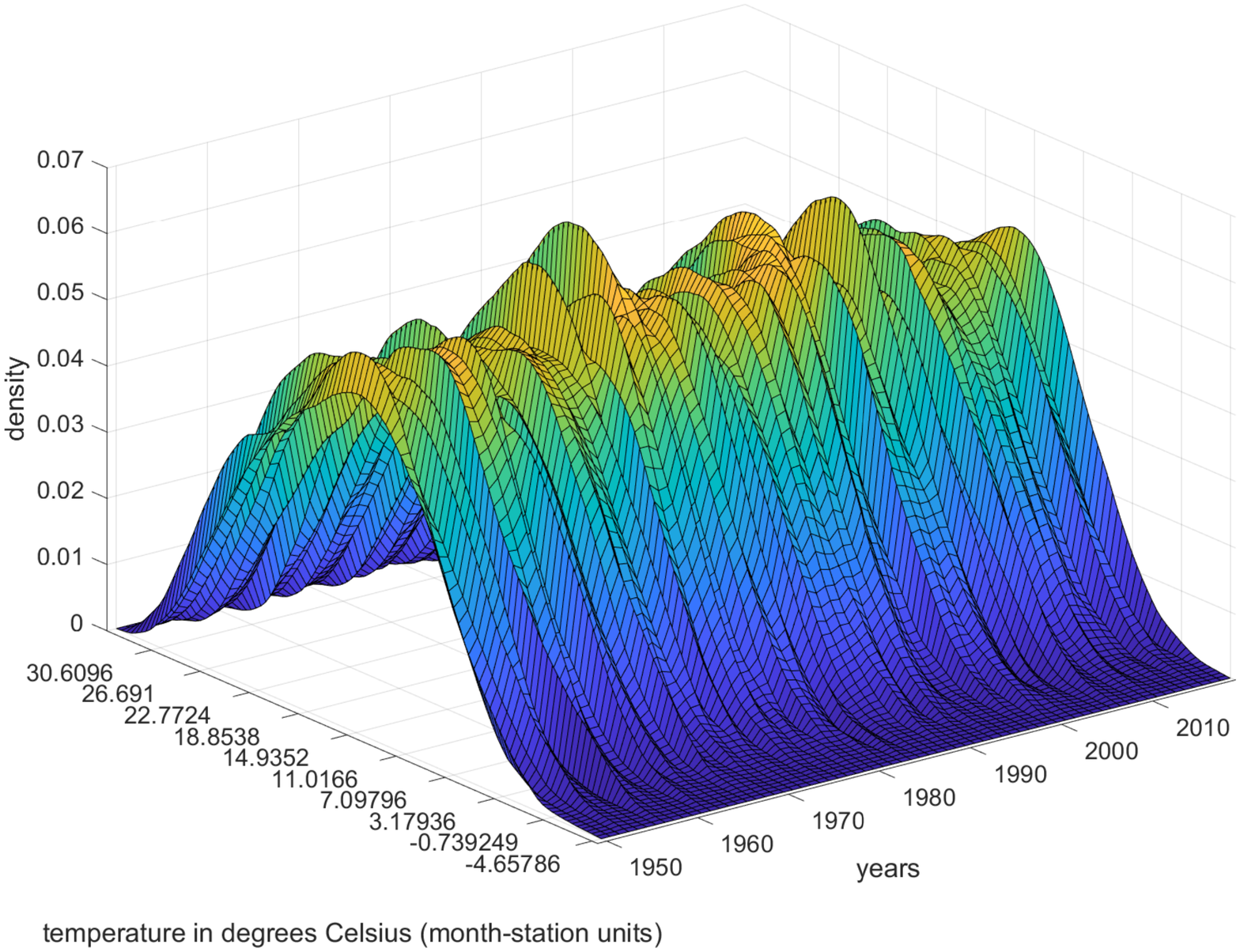}
	\end{center}
\end{figure}

\begin{figure}[h!]
	\begin{center}
		\caption{Characteristics of temperature data in Spain with stations selected since 1950 (monthly data across stations, AEMET, 1950-2019)} \label{fig-char-1950-monthly}
		\includegraphics[scale=0.5]{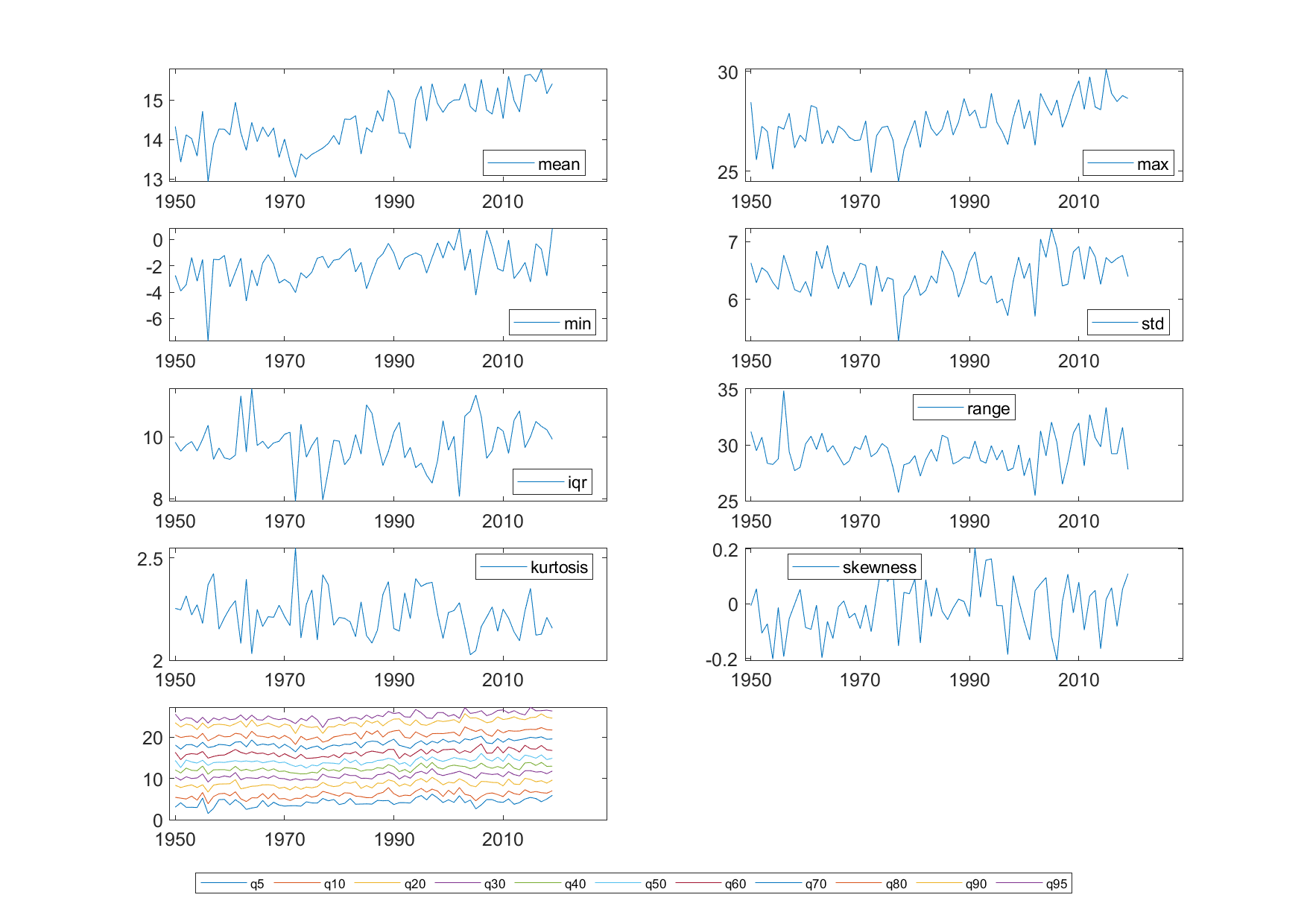}
	\end{center}
\end{figure}

The cross-sectional analysis is approached under two assumptions. First, choosing a sufficiently long and representative period of the geographical diversity of the Spanish Iberian Peninsula, 1950-2019. Second, we work with month-station units from daily observations to construct the annual observations of the time series object from the data supplied by the stations, following a methodology similar to that carried out for the whole planet in GG2020.\footnote{The results with daily averages are very similar. The decision to work with monthly data instead of daily in the cross-sectional approach has been based on its compatibility with the data available for the Globe. } The study comprises the steps described in the previous section. The density of the data and the evolution of characteristics are displayed, respectively in Figures \ref{fig-density-Spain} and \ref{fig-char-1950-monthly}.

We find positive and significant trends in the \textit{mean}, \textit{max}, \textit{min} and all the quantiles. Therefore from definition \ref{def1}, we conclude there exists a clear local warming (see Table \ref{tab-1950-Spain-monthly-rec-acc}).

The recursive evolution for the periods 1950-2019 and 1970-2019 shows a clear increase in the trends of the \textit{mean}, some dispersion measures and higher quantiles (see the last column of Table \ref{tab-1950-Spain-monthly-rec-acc}). More precisely, there is a significant trend acceleration in most of the distributional characteristics except the lower quantiles (below \textit{q20}). These quantiles, \textit{q05} and \textit{q10}, remain stable.
	
The co-trending tests for the full sample 1950-2019 show a similar evolution of the trend for all the quantiles with a constant \textit{iqr} (see Table  \ref{Tab-cotrend-since1950-monthly-1950}). This indicates that in this period the warming process of Spain can be considered a \textit{W1} type. More recently, 1970-2019, the co-trending tests (see Table \ref{Tab-cotrend-since1950-monthly-1970}) indicate the upper quantiles grow faster than the lower ones. This, together with a positive trend in the dispersion measured by the \textit{iqr} shows that Spain has evolved from a \textit{W1} to a \textit{W3} warming type process

Finally, no evidence of ``inner'' amplification during the period 1950-2019 is found in the lower quantiles. Regarding the upper quantiles, we found both ``inner'' and ``outer'' amplification in the second period, which supports the previous finding of a transition from type \textit{W1} to type \textit{W3} (see Table \ref{tab-amplif-Spain}).

Summing up, with our proposed tests for the evolution of the trend of the whole temperature distribution, we conclude that Spain has evolved from a \textit{W1} type to a much more dangerous \textit{W3} type. The results of acceleration and dynamic amplification reinforce the finding of this transition to type \textit{W3}.

	\begin{table}[h!]\caption{Trend acceleration hypothesis (Spain monthly data across stations, AEMET, 1950-2019)}\label{tab-1950-Spain-monthly-rec-acc}\begin{center}\scalebox{0.5}{\begin{tabular}{l*{5}{c}} \hline \hline
				& \multicolumn{2}{c}{Trend test by periods}& \multicolumn{1}{c}{Acceleration test}\\
				names/periods&1950-2019& 1970-2019& 1950-2019, 1970-2019\\ \hline
				mean  &0.0242&0.0389&3.0294\\
				& (0.0000)& (0.0000)& (0.0015) \\
				max   &0.0312&0.0526&2.7871\\
				& (0.0000)& (0.0000)& (0.0030) \\
				min   &0.0289&0.0251&-0.2557\\
				& (0.0000)& (0.0654)& (0.6007) \\
				std   &0.0036&0.0098&1.7952\\
				& (0.0518)& (0.0021)& (0.0374) \\
				iqr   &0.0051&0.0158&1.8197\\
				& (0.1793)& (0.0028)& (0.0355) \\
				rank  &0.0023&0.0276&1.2705\\
				& (0.8249)& (0.1127)& (0.1030) \\
				kur   &-0.0010&-0.0018&-0.9191\\
				& (0.0203)& (0.0198)& (0.8202) \\
				skw   &0.0011&-0.0002&-1.5989\\
				& (0.0271)& (0.7423)& (0.9439) \\
				q5    &0.0227&0.0206&-0.2559\\
				& (0.0000)& (0.0059)& (0.6008) \\
				q10   &0.0200&0.0203&0.0406\\
				& (0.0000)& (0.0077)& (0.4838) \\
				q20   &0.0209&0.0300&1.4158\\
				& (0.0000)& (0.0000)& (0.0796) \\
				q30   &0.0221&0.0333&2.0100\\
				& (0.0000)& (0.0000)& (0.0232) \\
				q40   &0.0213&0.0366&2.4867\\
				& (0.0000)& (0.0000)& (0.0071) \\
				q50   &0.0211&0.0404&3.2496\\
				& (0.0000)& (0.0000)& (0.0007) \\
				q60   &0.0246&0.0446&3.1147\\
				& (0.0000)& (0.0000)& (0.0011) \\
				q70   &0.0273&0.0478&3.3143\\
				& (0.0000)& (0.0000)& (0.0006) \\
				q80   &0.0275&0.0471&2.6949\\
				& (0.0000)& (0.0000)& (0.0040) \\
				q90   &0.0321&0.0548&3.2441\\
				& (0.0000)& (0.0000)& (0.0007) \\
				q95   &0.0335&0.0526&3.3568\\
				& (0.0000)& (0.0000)& (0.0005) \\
				\hline \hline \end{tabular}}\end{center}
	\begin{tablenotes}
		\tiny{		\textit{Note}: OLS estimates and HAC p-values in parenthesis of the $t_{\beta=0}$ test from regression: $C_{t}=\alpha+\beta t+u_{t}$, for two different time periods. For the acceleration hypothesis we run the system: $C_{t}=\alpha_{1} +\beta_{1} t+u_{t}, \text{   } t=1, ...,s ,..., T, C_{t}=\alpha_{2} +\beta_{2} t+u_{t}, \text{   } t=s+1, ..., T, \text{and test the null hypothesis } \beta_{2}=\beta_{1} \text{ against the alternative} \beta_{2}>\beta_{1}$. We show the value of the t-statistic and its HAC p-value.}
\end{tablenotes}\end{table}

\begin{table}[h!]\caption{Co-trending analysis (Spain monthly data across stations, AEMET, 1950-2019)}\label{Tab-cotrend-since1950-monthly-1950}\begin{center}\scalebox{0.7}{\begin{tabular}{l*{3}{c}} \hline \hline
				Joint hypothesis tests&Wald test&p-value\\ \hline
				All quantiles (q05, q10,...,q90, q95)&13.235&0.211 \\
				Lower quantiles (q05, q10, q20, q30)           &0.310&0.958 \\
				Medium quantiles (q40, q50, q60)         &0.438&0.803 \\
				Upper quantiles (q70, q80, q90, q95)          &1.515&0.679 \\
				Lower-Medium quantiles (q05, q10, q20, q30, q40, q50, q60)          &0.771&0.993 \\
				Medium-Upper quantiles (q40, q50, q60, q70, q80, q90, q95)          &8.331&0.215 \\
				Lower-Upper quantiles (q05, q10, q20,q30, q70, q80, q90, q95 )          &11.705&0.111 \\
				\hline
				Spacing hypothesis&Trend-coeff.&p-value\\ \hline
				q50-q05                            &-0.002&0.786 \\
				q95-q50&0.012&0.000 \\
				q95-q05                            &0.011&0.096 \\
				q75-q25  (iqr)                          &0.005&0.179 \\
				\hline \hline \end{tabular}}\end{center}
	\begin{tablenotes}
		\textit{Note}:   Annual distributional characteristics (quantiles) of temperature. The top panel shows the Wald test of the null hypothesis of equality
		of trend coefficients for a given set of characteristics. In the bottom panel, the TT is applied to the difference between two
		representative quantiles.
\end{tablenotes}\end{table}

\begin{table}[h!]\caption{Co-trending analysis (Spain monthly data across stations, AEMET, 1970-2019)}\label{Tab-cotrend-since1950-monthly-1970}\begin{center}\scalebox{0.7}{\begin{tabular}{l*{3}{c}} \hline \hline
				Joint hypothesis tests&Wald test&p-value\\ \hline
				All quantiles (q05, q10,...,q90, q95)&38.879&0.000 \\
				Lower quantiles (q05, q10, q20, q30)           &3.121&0.373 \\
				Medium quantiles (q40, q50, q60)         &1.314&0.518 \\
				Upper quantiles (q70, q80, q90, q95)          &1.719&0.633 \\
				Lower-Medium quantiles (q05, q10, q20, q30, q40, q50, q60)          &12.771&0.047 \\
				Medium-Upper quantiles (q40, q50, q60, q70, q80, q90, q95)          &10.675&0.099 \\
				Lower-Upper quantiles (q05, q10, q20,q30, q70, q80, q90, q95 )          &37.892&0.000 \\
				\hline
				Spacing hypothesis&Trend-coeff.&p-value\\ \hline
				q50-q05                            &0.020&0.029 \\
				q95-q50&0.012&0.050 \\
				q55-q05                            &0.032&0.002 \\
				q75-q25  (iqr)                          &0.016&0.003 \\
				\hline \hline \end{tabular}}\end{center}
	\begin{tablenotes}
		\textit{Note}:   Annual distributional characteristics (quantiles) of temperature. The top panel shows the Wald test of the null hypothesis of equality
		of trend coefficients for a given set of characteristics. In the bottom panel, the TT is applied to the difference between two
		representative quantiles.
\end{tablenotes}\end{table}

\begin{table}[h!]\caption{Amplification hypothesis (Spain monthly data, AEMET 1950-2019}\label{tab-amplif-Spain}\begin{center}\scalebox{0.8}{\begin{tabular}{l*{5}{c}} \hline \hline
				periods/variables&1950-2019&1970-2019&1950-2019&1970-2019\\ \hline
				& \multicolumn{2}{c}{Inner}& \multicolumn{2}{c}{Outer}\\  \hline
				q05&0.80&0.56&0.55&0.39\\
				& (0.866)& (0.998)& (0.990)& (0.996) \\
				q10&0.83&0.65&0.62&0.52\\
				& (0.899)& (0.994)& (0.992)& (0.986) \\
				q20&0.94&0.90&0.76&0.81\\
				& (0.816)& (0.890)& (0.993)& (0.899) \\
				q30&0.93&0.91&0.77&0.87\\
				& (0.935)& (0.929)& (0.997)& (0.834) \\
				q40&0.97&1.03&0.80&0.97\\
				& (0.744)& (0.318)& (0.978)& (0.566) \\
				q50&0.98&1.10&0.83&1.12\\
				& (0.612)& (0.067)& (0.944)& (0.212) \\
				q60&1.09&1.15&0.96&1.23\\
				& (0.103)& (0.051)& (0.619)& (0.056) \\
				q70&1.11&1.16&1.05&1.30\\
				& (0.040)& (0.006)& (0.350)& (0.028) \\
				q80&1.11&1.14&1.06&1.29\\
				& (0.083)& (0.071)& (0.325)& (0.060) \\
				q90&1.14&1.16&1.19&1.45\\
				& (0.101)& (0.118)& (0.078)& (0.007) \\
				q95&1.10&1.09&1.18&1.36\\
				& (0.089)& (0.191)& (0.051)& (0.008) \\
					\hline \hline \end{tabular}}\end{center}
			\begin{tablenotes}
		\textit{Note}: OLS estimates and HAC p-values of the t-statistic of testing $H_{0}: \beta_{i}=1$ versus $H_{a}: \beta_{i}>1$ in the regression: $C_{it}=\beta _{i0}+\beta _{i1} mean_{t}+\epsilon_{it}$. $mean$ refers to the average of the Spanish Global temperature distribution for the ``inner'' and ``outer''cases, respectively.
	\end{tablenotes}\end{table}

\clearpage
\subsection{Global warming: the Globe}
\begin{figure}[h!]
	\begin{center}
		\caption{Global annual temperature density calculated with monthly data across stations} \label{fig-density-Globe}
		\includegraphics[scale=0.5]{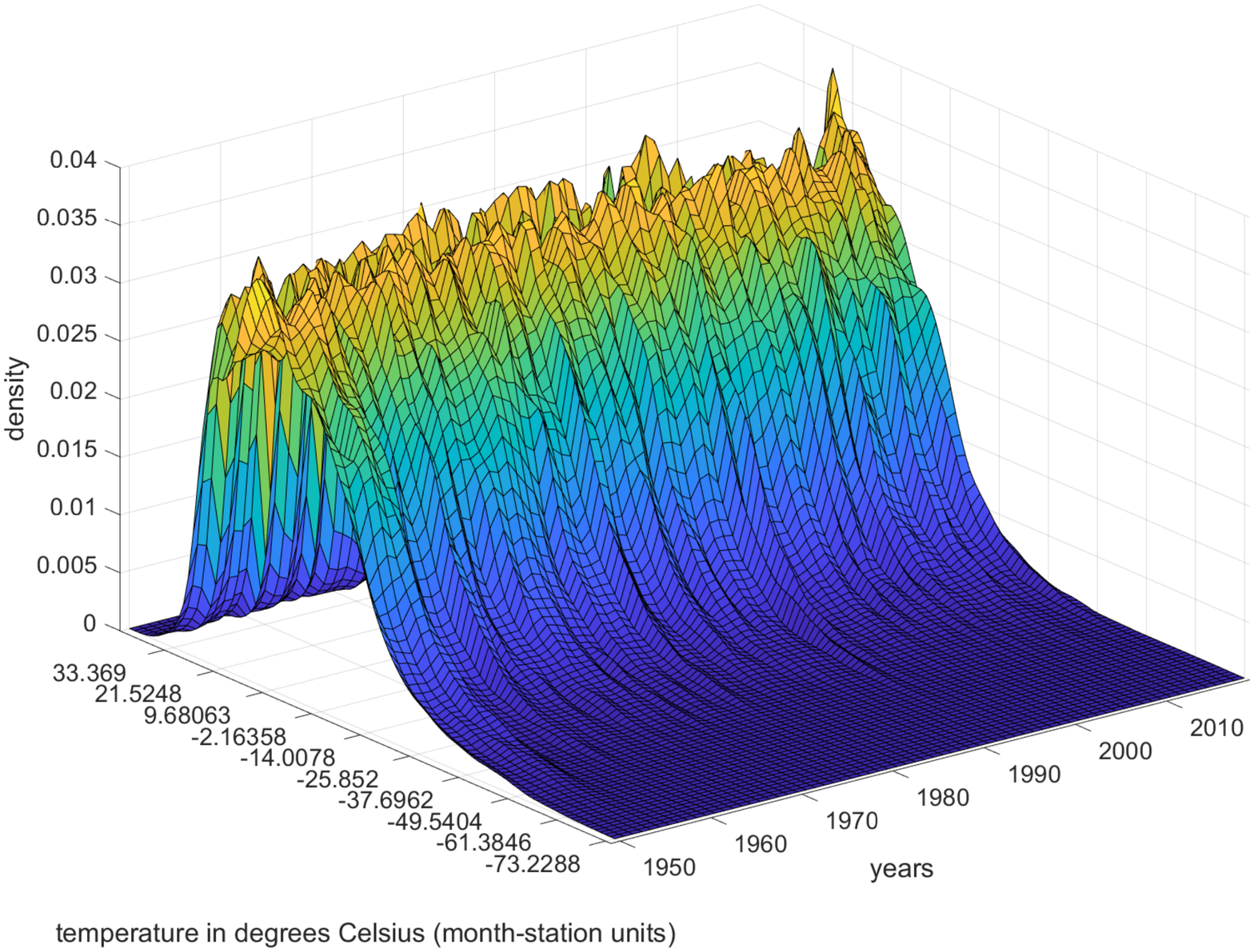}
	\end{center}
\end{figure}

\begin{figure}[h!]
	\begin{center}
		\caption{Characteristics of temperature data in the Globe (monthly data across stations, CRU, 1950-2019)} \label{fig-quantiles-Globe-monthly}
		\includegraphics[scale=0.5]{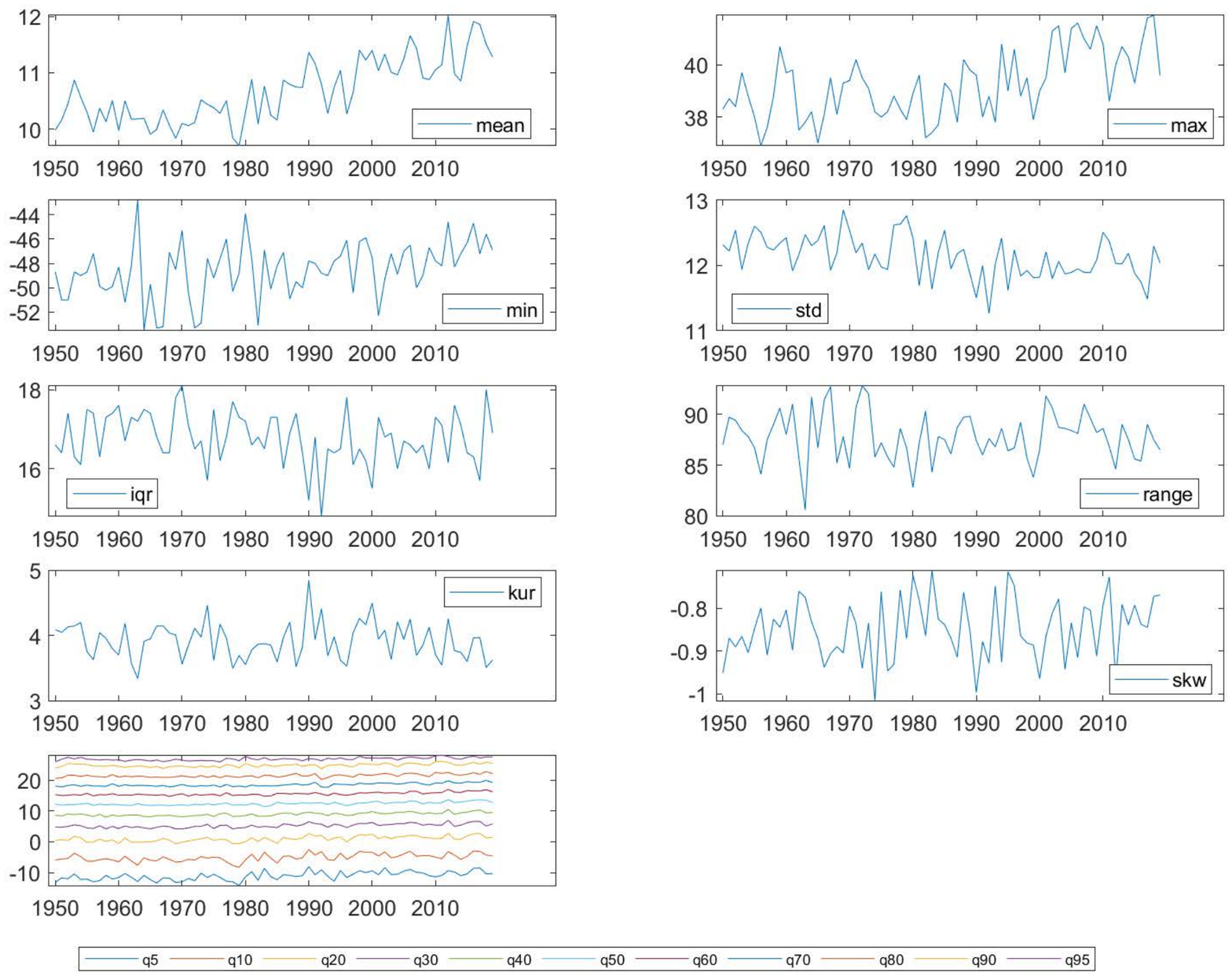}
	\end{center}
\end{figure}

In this section, we carry out a similar analysis to that described in the previous subsection for Spain.  Figures \ref{fig-density-Globe} and \ref{fig-quantiles-Globe-monthly} show the time evolution of the Global temperature densities and their different distributional characteristics from 1950 to 2019. The data in both figures are obtained from stations that report data throughout the sample period.

Table \ref{Tab-1950-Globe-monthly-acc} shows a positive trend in the mean as well as in all the quantiles. This indicates the clear existence of Global warming, more pronounced (larger trend) in the lower part of the distribution (a negative trend in the dispersion measures). The warming process suffers an acceleration in all the quantiles above \textit{q30}.

From the co-trending analysis (see Tables \ref{Tab-cotrend-Globe-monthly-1950-2019} and \ref{Tab-cotrend-Globe-monthly-1970-2019}) we can determine the type of warming process characterizing the whole Globe. Table \ref{Tab-cotrend-Globe-monthly-1950-2019} indicates that in the period 1950-2019 the Globe experimented a \textit{W2} warming type (the lower part of the temperature distribution grows faster than the middle and upper part, implying \textit{iqr} and \textit{std} have a negative trend).  Similar results are maintained for the period 1970-2019 (in this case only the dispersion measure \textit{std} has a negative trend).

The asymmetric amplification results shown in Table \ref{tab-amplif-Globe} reinforce the \textit{W2} typology for the whole Globe: an increase of one degree in the global mean temperature increases the lower quantiles by more than one degree. This does not occur with the upper part of the distribution. Notice that this amplification goes beyond the standard Artic amplification (\textit{q05}) affecting also \textit{q10}, \textit{q20} and \textit{q30}.

Summing up, the results from our different proposed tests for the evolution of the trend of the whole temperature distribution indicate that the Globe can be cataloged as a undergoing type \textit{W2} warming process. This warming type may have more serious consequences for ice melting, sea level increases, permafrost, $CO_{2}$ migration, etc. than the other types.

	\begin{table}[h!]\caption{Trend acceleration hypothesis (CRU monthly data across stations, 1950-2019)}\label{Tab-1950-Globe-monthly-acc}\begin{center}\scalebox{0.5}{\begin{tabular}{l*{5}{c}} \hline \hline
				& \multicolumn{2}{c}{Trend test by periods}& \multicolumn{1}{c}{Acceleration test}\\
				names/periods&1950-2019& 1970-2019& 1950-2019, 1970-2019\\ \hline
				mean  &0.0213&0.0300&2.2023\\
				& (0.0000)& (0.0000)& (0.0147) \\
				max   &0.0361&0.0523&1.1217\\
				& (0.0000)& (0.0001)& (0.1320) \\
				min   &0.0423&-0.0109&0.5016\\
				& (0.0000)& (0.5867)& (0.3084) \\
				std   &-0.0070&-0.0057&0.1776\\
				& (0.0000)& (0.0570)& (0.4296) \\
				iqr   &-0.0067&-0.0043&0.2454\\
				& (0.0435)& (0.4183)& (0.4033) \\
				rank  &-0.0062&0.0632&0.2181\\
				& (0.5876)& (0.0005)& (0.4138) \\
				kur   &-0.0010&0.0001&0.0445\\
				& (0.5205)& (0.9566)& (0.4823) \\
				skw   &0.0006&0.0003&0.0301\\
				& (0.0577)& (0.5726)& (0.4880) \\
				q5    &0.0404&0.0468&0.7035\\
				& (0.0000)& (0.0000)& (0.2415) \\
				q10   &0.0305&0.0406&0.9273\\
				& (0.0000)& (0.0001)& (0.1777) \\
				q20   &0.0253&0.0342&1.0156\\
				& (0.0000)& (0.0000)& (0.1558) \\
				q30   &0.0215&0.0280&1.2056\\
				& (0.0000)& (0.0000)& (0.1150) \\
				q40   &0.0192&0.0293&1.9873\\
				& (0.0000)& (0.0000)& (0.0245) \\
				q50   &0.0179&0.0268&1.8614\\
				& (0.0000)& (0.0000)& (0.0324) \\
				q60   &0.0185&0.0291&2.1971\\
				& (0.0000)& (0.0000)& (0.0149) \\
				q70   &0.0185&0.0288&2.5770\\
				& (0.0000)& (0.0000)& (0.0055) \\
				q80   &0.0160&0.0257&2.2460\\
				& (0.0000)& (0.0000)& (0.0132) \\
				q90   &0.0146&0.0243&2.0848\\
				& (0.0005)& (0.0000)& (0.0195) \\
				q95&0.0143&0.0239&1.7520\\
				& (0.0001)& (0.0000)& (0.0410) \\
				\hline \hline \end{tabular}}\end{center}
	\begin{tablenotes}
		\tiny{		\textit{Note}: OLS estimates and HAC p-values in parenthesis of the $t_{\beta=0}$ test from regression: $C_{t}=\alpha+\beta t+u_{t}$, for two different time periods. For the acceleration hypothesis we run the system: $C_{t}=\alpha_{1} +\beta_{1} t+u_{t}, \text{   } t=1, ...,s ,..., T, C_{t}=\alpha_{2} +\beta_{2} t+u_{t}, \text{   } t=s+1, ..., T, \text{and test the null hypothesis } \beta_{2}=\beta_{1} \text{ against the alternative} \beta_{2}>\beta_{1}$. We show the value of the t-statistic and its HAC p-value.}
\end{tablenotes}\end{table}

\begin{table}[h!]\caption{Co-trending analysis (CRU montly data, 1950-2019)}\label{Tab-cotrend-Globe-monthly-1950-2019}\begin{center}\scalebox{0.7}{\begin{tabular}{l*{3}{c}} \hline \hline
				Joint hypothesis tests&Wald test&p-value\\ \hline
				All quantiles (q05, q10,...,q90, q95)&25.143&0.005 \\
				Lower quantiles (q05, q10, q20, q30)           &9.545&0.023 \\
				Medium quantiles (q40, q50, q60)         &0.078&0.962 \\
				Upper quantiles (q70, q80, q90, q95)          &1.099&0.777 \\
				Lower-Medium quantiles (q05, q10, q20, q30, q40, q50, q60)          &17.691&0.007 \\
				Medium-Upper quantiles (q40, q50, q60, q70, q80, q90, q95)          &2.041&0.916 \\
				Lower-Upper quantiles (q05, q10, q20,q30, q70, q80, q90, q95 )          &24.683&0.001 \\
				\hline
				Spacing hypothesis&Trend-coeff.&p-value\\ \hline
				q50-q05                            &-0.022&0.000 \\
				q95-q50&-0.004&0.193 \\
				q95-q05                            &-0.026&0.000 \\
				q75-q25  (iqr)                          &-0.007&0.043 \\
					\hline \hline \end{tabular}}\end{center}
			\begin{tablenotes}
		\textit{Note}:   Annual distributional characteristics (quantiles) of temperature. The top panel shows the Wald test of the null hypothesis of equality
	of trend coefficients for a given set of characteristics. In the bottom panel, the TT is applied to the difference between two
representative quantiles.
\end{tablenotes}\end{table}

\begin{table}[h!]\caption{Co-trending analysis (CRU montly data, 1970-2019)}\label{Tab-cotrend-Globe-monthly-1970-2019}\begin{center}\scalebox{0.7}{\begin{tabular}{l*{3}{c}} \hline \hline
				Joint hypothesis tests&Wald test&p-value\\ \hline
				All quantiles (q05, q10,...,q90, q95)&18.478&0.047 \\
				Lower quantiles (q05, q10, q20, q30)           &5.523&0.137 \\
				Medium quantiles (q40, q50, q60)         &0.569&0.752 \\
				Upper quantiles (q70, q80, q90, q95)          &2.667&0.446 \\
				Lower-Medium quantiles (q05, q10, q20, q30, q40, q50, q60)          &7.606&0.268 \\
				Medium-Upper quantiles (q40, q50, q60, q70, q80, q90, q95)          &6.714&0.348 \\
				Lower-Upper quantiles (q05, q10, q20,q30, q70, q80, q90, q95 )          &14.520&0.043 \\
				\hline
				Spacing hypothesis&Trend-coeff.&p-value\\ \hline
				q50-q05                            &-0.020&0.047 \\
				q95-q50&-0.003&0.462 \\
				q95-q05                            &-0.023&0.048 \\
				q75-q25  (iqr)                          &-0.004&0.418 \\
					\hline \hline \end{tabular}}\end{center}
			\begin{tablenotes}
		\textit{Note}:  Annual distributional characteristics (quantiles) of temperature. The top panel shows the Wald test of the null hypothesis of equality
	of trend coefficients for a given set of characteristics. In the bottom panel, the TT is applied to the difference between two
representative quantiles.
\end{tablenotes}\end{table}

			\begin{table}[h!]\caption{Amplification hypotheses (CRU monthly data across stations, 1950-2019)}\label{tab-amplif-Globe}\begin{center}\scalebox{0.8}{\begin{tabular}{l*{3}{c}} \hline \hline
							periods/variables&1950-2019&1970-2019\\ \hline
					q05&2.00&1.83\\
					& (0.000)& (0.000) \\
					q10&1.79&1.73\\
					& (0.000)& (0.001) \\
					q20&1.41&1.37\\
					& (0.000)& (0.000) \\
					q30&1.07&1.00\\
					& (0.089)& (0.502) \\
					q40&0.88&0.91\\
					& (0.999)& (0.973) \\
					q50&0.74&0.81\\
					& (1.000)& (0.997) \\
					q60&0.74&0.85\\
					& (0.999)& (0.973) \\
					q70&0.77&0.85\\
					& (1.000)& (0.988) \\
					q80&0.72&0.78\\
					& (1.000)& (1.000) \\
					q90&0.69&0.70\\
					& (1.000)& (1.000) \\
					q95&0.60&0.64\\
					& (1.000)& (1.000) \\ 				
													\hline \hline \end{tabular}}\end{center}
							\begin{tablenotes}
						\textit{Note}: OLS estimates and HAC p-values of the t-statistic of testing $H_{0}: \beta_{i}=1$ versus $H_{a}: \beta_{i}>1$ in the regression: $C_{it}=\beta _{i0}+\beta _{i1} mean_{t}+\epsilon_{it}$. $mean$ refers to the average of the Global temperature distribution.
					\end{tablenotes}\end{table}
\clearpage			
\subsection{Micro-local warming: Madrid and Barcelona}
The existence of warming heterogeneity implies that in order to design more efficient mitigation policies, they have to be developed at different levels: global, country, region etc. How local we need to go will depend on the existing degree of micro-warming heterogeneity. In this subsection, we go to the smallest level, climate station level . We analyze, within Spain, the warming process in two weather stations corresponding to two cities: Madrid (Retiro station) and Barcelona (Fabra station). \footnote{From Madrid and Barcelona there is data since 1920's, nevertheless we began the study in 1950 for consistency with the previous analysis of Spain and the Globe.}  Obviously, the data provided by these stations is not cross-sectional data but directly pure time series data. Our methodology can be easily applied to higher frequency time series, in this case daily data, to compute the distributional characteristics (see Figures \ref{fig-char-daily-Madrid-1950} and \ref{fig-char-daily-Barcelona-1950})\footnote{See the application to Central England in GG2020 and in Gadea and Gonzalo (2022) to Madrid, Zaragoza and Oxford.}.

The results are shown in the Appendix. These two stations, Madrid-Retiro and Barcelona-Fabra clearly experience two different types of warming. First, there is evidence of micro-local warming, understood as the presence of significant and positive trends, in all the important temperature distributional characteristics of both stations. The acceleration phenomenon is also clearly detected, in other words, the warming increases as time passes (see Tables \ref{Tab-1950-Madrid-daily-rec-acc}  and  \ref{Tab-1950-Barcelona-daily-rec-acc}). Secondly, from the co-trending tests (Tables \ref{Tab-cotrend-Madrid-daily-1950}-\ref{Tab-cotrend-Madrid-daily-1970} and \ref{Tab-cotrend-Barcelona-daily-1950}-\ref{Tab-cotrend-Barcelona-daily-1970}), it can be concluded that the warming process of Madrid-Retiro is type \textit{W3} while for Barcelona-Fabra  it is type \textit{W1}. In both cases the warming typology is stable  through both sample periods (1950-2019 and 1970-2019). Thirdly, as expected, Madrid-Retiro presents ``inner'' and  ``outer'' amplification for the upper quantiles, while Barcelona-Fabra does so only for the center part of its temperature distribution (see Tables \ref{Tab-amplif-Madrid-1950} and \ref{Tab-amplif-Barcelona-1950}).

Summing up, even within Spain we find evidence of warming heterogeneity. While Madrid (Continental Mediterranean climate) has a similar pattern as that of peninsular Spain (1970-2019) \textit{W3}, Barcelona (Mediterranean coastline climate) maintains a \textit{W1} typology.  Thus there are two different warming processes which require mitigation policies at the country as well as the very local level.

\section{Comparing results}
\begin{figure}[h!]
	\begin{center}
		\caption{Trend evolution of different temperature distributional characteristics} \label{fig-comp-Globe-Spain-Madrid-Barcelona}
		\includegraphics[scale=0.5]{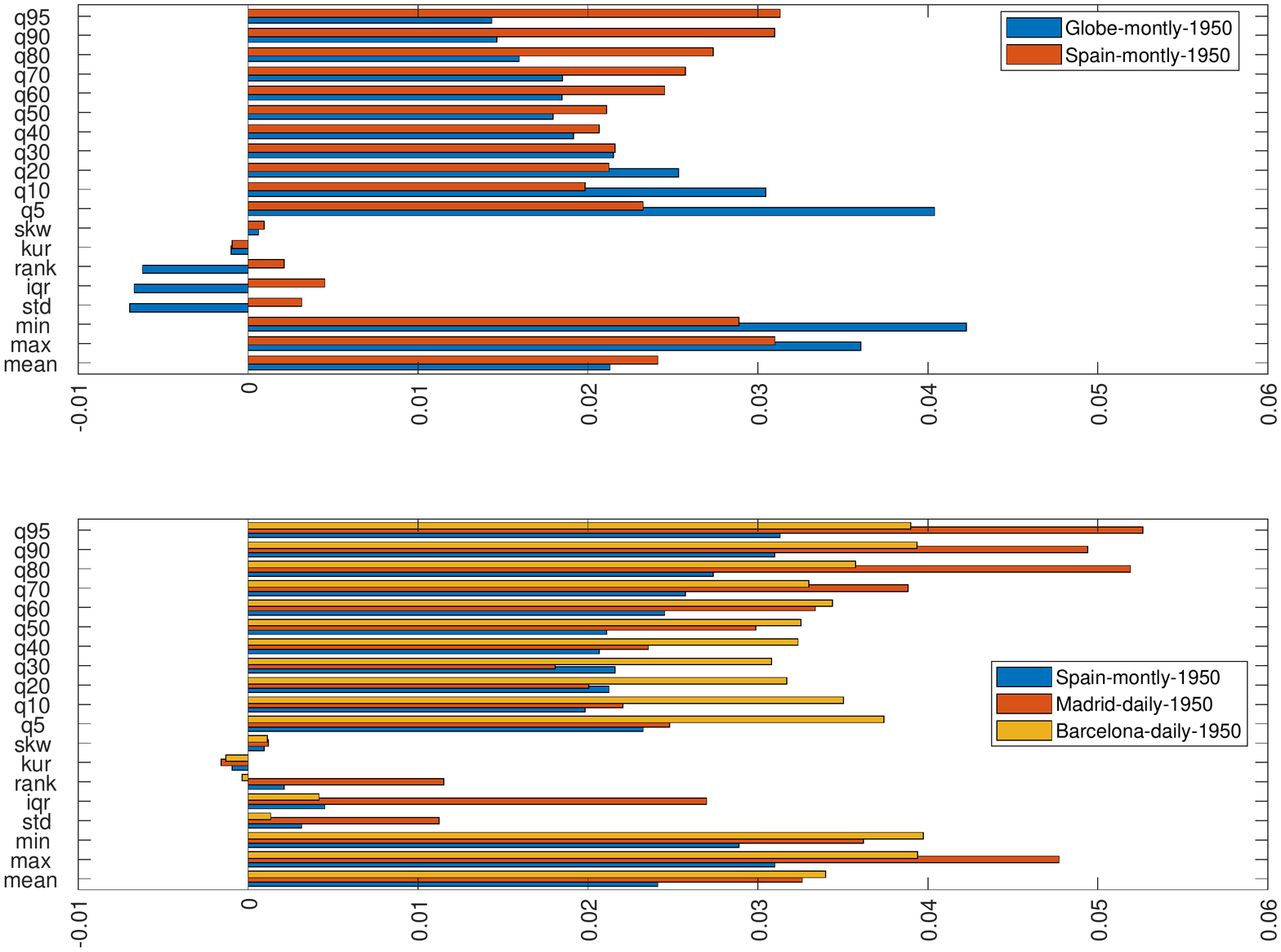}
	\end{center}
\begin{figurenotes}
\textit{Note}: The bars represent the intensity of the trends found in each characteristic measured through the value of the $\beta$-coefficient estimated in the regression $C_{t}=\alpha+\beta t+u_{t}$.
\end{figurenotes}
\end{figure}

The goal of this section is to show the existence of climate heterogeneity by comparing the results obtained from applying our three-step methodology to different regions. These results are summarized in Table \ref{Tab-summary}. It is clear that there is distributional warming in all the analyzed areas; but this warming follows different patterns and sometimes the warming type is not even stable. In the case of Spain, it depends on the period under consideration. Figure \ref{fig-comp-Globe-Spain-Madrid-Barcelona} captures graphically the different trend behavior and intensity of the distributional characteristics by regions (Spain and the Globe and Madrid and Barcelona).\footnote{The analysis of other characteristics such as the third and fourth order moments can contribute to the temperature distributions. In the case of Spain, the kurtosis is always negative with a mean value of -0.8 and a significant negative trend, which means that we are dealing with a platykurtic distribution with tails less thick than Normal, a shape that is accelerating over time. However, it is ot possible to draw conclusions about symmetry given its high variability over time. Conversely, the temperature distribution in the Globe is clearly leptokurtic with an average kurtosis of 0.9 and a negative but not significant trend. The global temperature observations are therefore more concentrated around the mean and their tails are thicker than in a Normal distribution. The skewness is clearly negative although a decreasing and significant trend points to a reduction of the negative skewness. } The graphical results in this figure coincide with the results of the warming typology tests shown in Table \ref{Tab-summary}.

\begin{figure}[h!]
	\begin{center}
		\caption{Comparing heatmaps}
		\label{fig-comp-Globe-Spain-heatmap}
		\subfloat[{\small Globe}]{
			\includegraphics[scale=0.4]{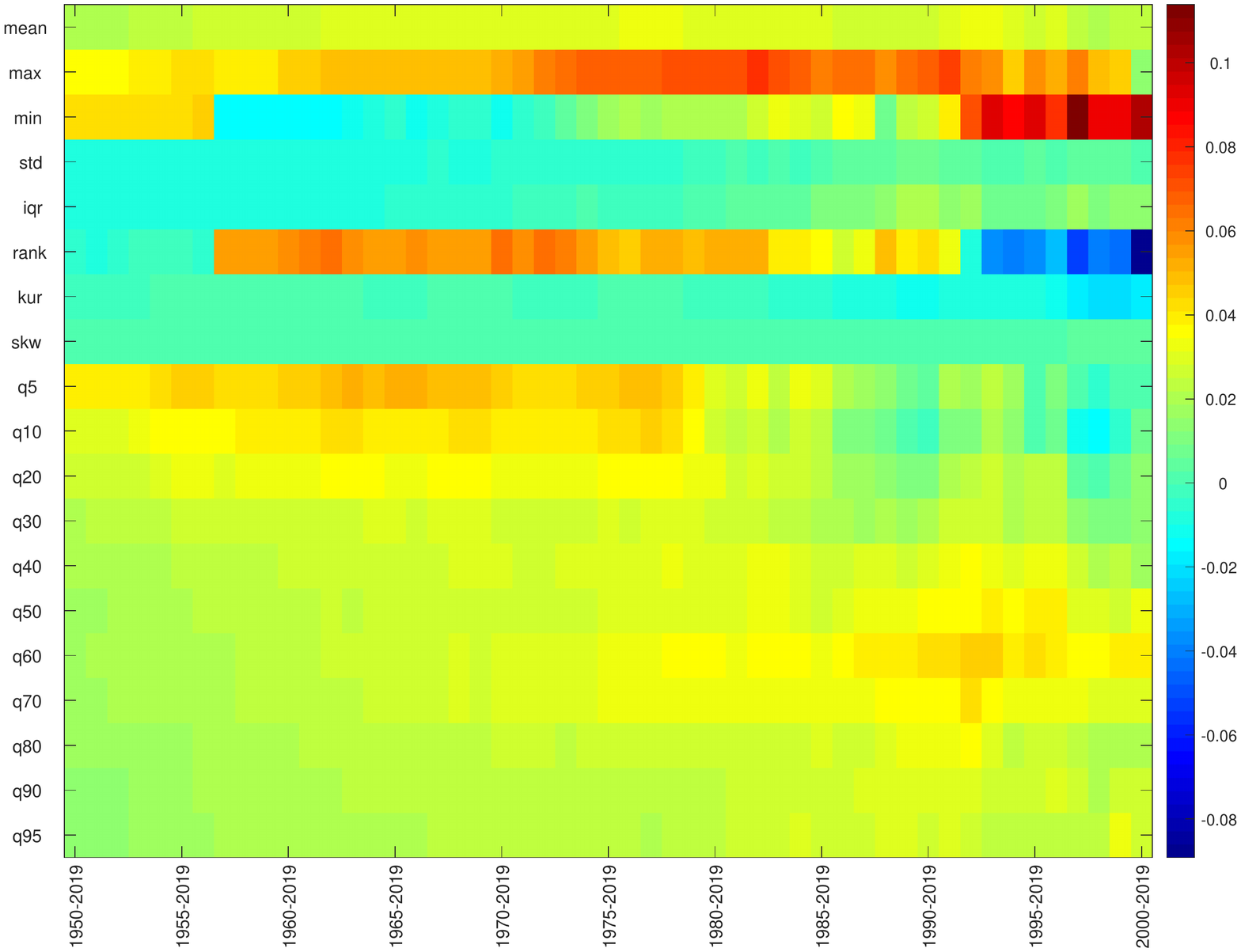}}\\
		\subfloat[{\small Spain}]{
			\includegraphics[scale=0.4]{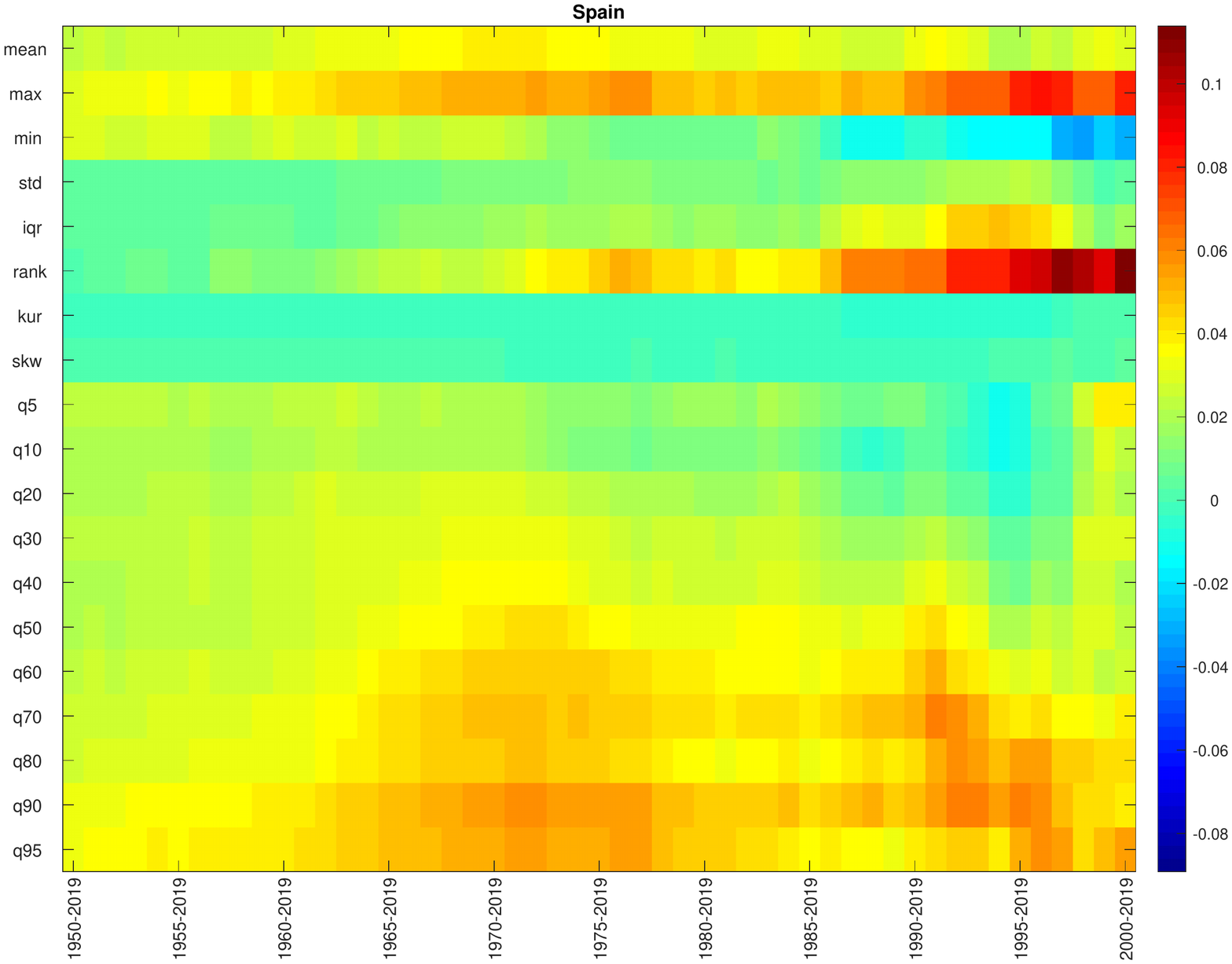}}\\
	\end{center}
	\begin{figurenotes}
		\textit{Note}: The color scale on the right side of the figure shows the intensity of the trend, based on  the value of the $\beta$-coefficient estimated in the regression $C_{t}=\alpha+\beta t+u_{t}$.
	\end{figurenotes}
\end{figure}

The middle of Table \ref{Tab-summary} shows that warming acceleration is detected in all the locations. This acceleration is more general in Spain than in the Globe (see also the heatmap in Figure \ref{fig-comp-Globe-Spain-heatmap}) and in Barcelona than in Madrid. Apart from these differences, the acceleration shares certain similarities across regions. This is not the case for the warming amplification that is clearly asymmetric. Spain suffers an amplification in the upper quantiles while the Globe does so in the lower ones. Notice that the latter amplification goes beyond the standard results found in the literature for the Arctic region (\textit{q05}). We detect amplification also for the regions corresponding to the quantiles \textit{q10}-\textit{q30}. In the case of Madrid and Barcelona, Madrid suffers a wider warming amplification than Barcelona.

The results of the first two steps of our methodology are obtained region by region (Spain, the Globe, Madrid and Barcelona). It is the last step, via the warming dominance test (see the numerical results in Table \ref{tab-WD}) where we compare directly one region with another. Warming in Spain dominates that of the Globe in all the quantiles except the lower \textit{q05}.\footnote{A more detailed analysis of the warming process suffered in the Artic region can be found in Gadea and Gonzalo (2021).} This would support the idea held in European institutions and gathered in international reports on the greater intensity of climate change in the Iberian Peninsula. Warming in Madrid dominates that of Barcelona in the upper quantiles, while the reverse is the case in the lower quantiles. This latter result coincides with the idea that regions close to the sea have milder upper temperatures.

Further research (beyond the scope of this paper) will go in the direction of finding the possible causes behind the warming types \textit{W1}, \textit{W2}, and \textit{W3}. Following the literature, on diurnal temperature asymmetry (Diurnal Temperature Range $=DTR= T_{max}-T_{min}$) we can suggest as  possible causes for \textit{W2} the cloud coverage (Karl et al. 1993) and the planetary boundary layer (see Davy et al. 2017). For \textit{W3}, the process of desertification  (see Karl et al. 1993).

Summarizing, in this section we describe, measure and test the existence of warming heterogeneity in different regions of the planet. It is important to note that these extensive results can not be obtained by the standard analysis of the average temperature.

\begin{table}[h!]\caption{Warming dominance}\label{tab-WD}\begin{center}\scalebox{1}{\begin{tabular}{lcccc} \hline \hline
				& \multicolumn{2}{c}{Spain-Globe}& \multicolumn{2}{c}{Madrid-Barcelona}\\
				Quantile&$\beta$&t-ratio&$\beta$&t-ratio\\ \hline
				q05    &-0.018&(-2.770)&-0.013&(-3.730)\\
				q10   &-0.010&(-1.504)&-0.013&(-4.215)\\
				q20   &-0.004&(-0.950)&-0.012&(-2.988)\\
				q30   &0.001&(0.180)&-0.013&(-4.164)\\
				q40   &0.002&(0.788)&-0.009&(-2.909)\\
				q50   &0.003&(1.025)&-0.003&(-0.701)\\
				q60   &0.006&(1.933)&-0.001&(-0.219)\\
				q70   &0.009&(3.266)&0.006&(1.252)\\
				q80   &0.012&(3.203)&0.016&(3.331)\\
				q90   &0.017&(3.862)&0.010&(1.869)\\
				q95   &0.019&(4.930)&0.014&(1.993)\\
				\hline \hline
			\end{tabular}
	}\end{center}
	\begin{tablenotes}
		\textit{Note}: The slopes (t-statistic) of the following regression \begin{equation*}
			q_{\tau t}(A)- q_{\tau t}(B)=\alpha_{\tau} +\beta_{\tau} t +u_{\tau t}
		\end{equation*}
		In the first column \textit{A}=Spain, \textit{B}=Globe and in the second \textit{A}=Madrid, \textit{B}=Barcelona.
	\end{tablenotes}
\end{table}

	\begin{table}[h!]
	\caption{Summary of results}\label{Tab-summary}\begin{center}\scalebox{0.65}{
			\begin{tabular}{c|c|c|c|c|c|c} \\ \hline \hline
				\multicolumn{7}{c}{Cross analysis} \\ \hline
				Sample & Period & Type & Acceleration & \multicolumn{2}{c}{Amplification} & Dominance \\ \hline
				&  &  &  & Inner & Outer &  \\
				Spain & &  &  &  &  & \\
				& 1950-2019 & \textit{W1} & [\textit{mean, std, iqr, rank, } & [\textit{q70, q80, q95]} & [\textit{q90, q95]} & [q60,..., q95] \\
				& & & \textit{q20,..., q95]} &  	&  \\
				& 1970-2019 & \textit{W3} &  & [\textit{q50,..., q80]} & [\textit{q60,..., q95]}  &   \\
				The Globe &  &  &  &  &  & \\
				& 1950-2019 & \textit{W2} & [\textit{mean} & [\textit{q05,..., q30]} & &  [\textit{q05]}  \\
				&  &  & \textit{q40,..., q95]} &  &  &  \\
				& 1970-2019 & \textit{W2} &  & [\textit{q05,..., q20]} & &  \\
				&  &  &  &  &  &  \\ \hline
				\multicolumn{7}{c}{Time analysis} \\ \hline
				Sample & Period & Type & Acceleration & \multicolumn{2}{c}{Amplification} & Dominance \\ \hline
				Madrid, Retiro Station &  & &  & &  & \\
				& 1950-2019 & \textit{W3} & [\textit{mean, std, rank,  } & [\textit{q50,..., q95]} & [\textit{	q40,..., q95]} & [q80,..., q95]   \\
				&  &  & \textit{q40, ..., q95]} &  &  &  \\
				& 1970-2019 & \textit{W3} &  & [\textit{q50,..., q95]} & [\textit{q40,..., q95]} &  \\
				Barcelona, Fabra Station  &  &  & &  &  &  \\
				& 1950-2019 & \textit{W1} & [\textit{mean, } & \textit{-} & [\textit{q30,..., q90]} & \textit{[q05,..., q40]} \\
				&  &  & \textit{q20,..., q95]} &  &  &  \\
				& 1970-2019 & \textit{W1} &  & [\textit{q60, q70]} & [\textit{q30,..., q70]} &  \\
				&  &  &  &  &  &  \\ \hline \hline
			\end{tabular}
	}\end{center}
	\begin{tablenotes}
		\tiny{		\textit{Note}: For Spain and the Globe we build characteristics from station-months units. For Madrid and Barcelona we use daily frequency time series. A significance level of 10\% is considered for all tests and characteristics.}
	\end{tablenotes}
\end{table}

\clearpage
\section{Conclusions}
The existence of Global Warming is very well documented in all the scientific reports published by the IPCC. In the last one, the AR6 report (2022), special attention is dedicated to climate change heterogeneity (regional climate). Our paper presents a new quantitative methodology, based on the evolution of the trend of the whole temperature distribution and not only on the average, to characterize, to measure and to test the existence of such warming heterogeneity.
It is found that the local warming experienced by Spain (one of most climatically diverse areas) is very different from that of the Globe as a whole. In Spain, the upper-temperature quantiles tend to increase more than the lower ones, while in the Globe just the opposite occurs. In both cases the  warming process is accelerating over time. Both regions suffer an amplification effect of an asymmetric nature: there is warming amplification in the lower quantiles of the Globe temperature (beyond the standard well-known results of the Arctic zone) and in the upper ones of Spain. Overall, warming in Spain dominates  that of the Globe in all the quantiles except the lower \textit{q05}. This places Spain in a very difficult warming situation compared to the Globe. Such a situation  requires stronger mitigation-adaptation policies. For this reason, future climate agreements should take into consideration the whole temperature distribution and not only the average.

Any time a novel methodology is proposed, new research issues emerge for future investigation. Among those which have been left out of this paper (some are part of our current research agenda), three points stand out as important:

\begin{itemize}
\item	There is a clear need for a new non-uniform causal-effect climate change analysis beyond the standard causality in mean.
\item	In order to improve efficiency, mitigation-adaptation policies should be designed containing a common global component and an idiosyncratic regional element.
\item	The relation between warming heterogeneity  and public awareness of climate change deserves to be analyzed. 	
\end{itemize}

\clearpage

\section{Appendix: Climate change of Madrid and Barcelona}	

\setcounter{table}{0}
\renewcommand{\thetable}{A\arabic{table}}

\setcounter{figure}{0}
\renewcommand{\thefigure}{A\arabic{figure}}

\subsection{Madrid-Retiro}
\begin{figure}[h!]
	\begin{center}
		\caption{Characteristics of temperature data in Madrid-Retiro (AEMET daily data, 1950-2019)} \label{fig-char-daily-Madrid-1950}
		\includegraphics[scale=0.55]{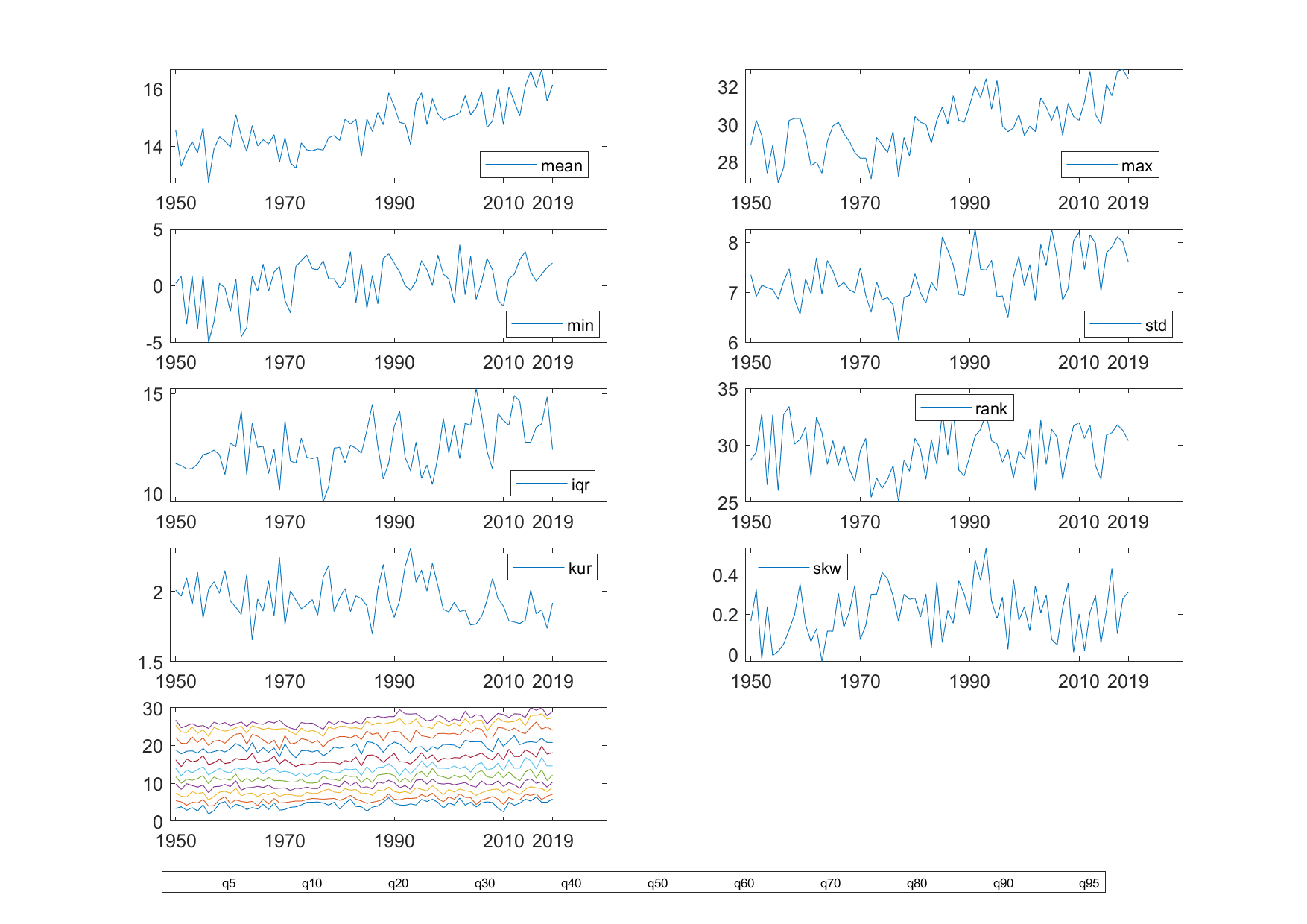}
	\end{center}
\end{figure}

\begin{table}\caption{Trend acceleration hypothesis (Madrid, daily data, AEMET, 1950-2019)}\label{Tab-1950-Madrid-daily-rec-acc}\begin{center}\scalebox{0.5}{\begin{tabular}{l*{5}{c}} \hline \hline
				& \multicolumn{2}{c}{Trend test by periods}& \multicolumn{1}{c}{Acceleration test}\\
				names/periods&1950-2019& 1970-2019& 1950-2019, 1970-2019\\ \hline
				mean  &0.0326&0.0447&2.0972\\
				& (0.0000)& (0.0000)& (0.0189) \\
				max   &0.0477&0.0636&1.2043\\
				& (0.0000)& (0.0000)& (0.1153) \\
				min   &0.0362&0.0087&-1.5077\\
				& (0.0011)& (0.5859)& (0.9330) \\
				std   &0.0112&0.0197&2.1160\\
				& (0.0000)& (0.0000)& (0.0181) \\
				iqr   &0.0270&0.0399&1.1110\\
				& (0.0000)& (0.0004)& (0.1343) \\
				rank  &0.0115&0.0549&2.0160\\
				& (0.3666)& (0.0045)& (0.0229) \\
				kur   &-0.0016&-0.0022&-0.4449\\
				& (0.0278)& (0.0660)& (0.6714) \\
				skw   &0.0012&-0.0013&-1.7769\\
				& (0.1538)& (0.2695)& (0.9611) \\
				q5    &0.0248&0.0183&-0.5712\\
				& (0.0000)& (0.0774)& (0.7156) \\
				q10   &0.0220&0.0174&-0.5815\\
				& (0.0000)& (0.0162)& (0.7191) \\
				q20   &0.0200&0.0187&-0.1777\\
				& (0.0000)& (0.0099)& (0.5704) \\
				q30   &0.0181&0.0235&0.6959\\
				& (0.0000)& (0.0019)& (0.2438) \\
				q40   &0.0236&0.0362&1.6625\\
				& (0.0000)& (0.0000)& (0.0494) \\
				q50   &0.0299&0.0545&2.8801\\
				& (0.0000)& (0.0000)& (0.0023) \\
				q60   &0.0334&0.0604&3.1655\\
				& (0.0000)& (0.0000)& (0.0010) \\
				q70   &0.0388&0.0550&1.7385\\
				& (0.0000)& (0.0000)& (0.0422) \\
				q80   &0.0519&0.0712&1.9750\\
				& (0.0000)& (0.0000)& (0.0251) \\
				q90   &0.0494&0.0687&1.7956\\
				& (0.0000)& (0.0000)& (0.0374) \\
				q95   &0.0527&0.0710&1.7839\\
				& (0.0000)& (0.0000)& (0.0383) \\
				\hline \hline \end{tabular}}\end{center}
	\begin{tablenotes}
		\tiny{		\textit{Note}: OLS estimates and HAC p-values in parenthesis of the $t_{\beta=0}$ test from regression: $C_{t}=\alpha+\beta t+u_{t}$, for two different time periods. For the acceleration hypothesis we run the system: $C_{t}=\alpha_{1} +\beta_{1} t+u_{t}, \text{   } t=1, ...,s ,..., T, C_{t}=\alpha_{2} +\beta_{2} t+u_{t}, \text{   } t=s+1, ..., T, \text{and test the null hypothesis } \beta_{2}=\beta{1} \text{ against the alternative} \beta_{2}>\beta{1}$. We show the value of the t-statistic and its HAC p-value.}
\end{tablenotes}\end{table}

\begin{table}\caption{Co-trending analysis (Madrid-Retiro daily data, AEMET 1950-2019)}\label{Tab-cotrend-Madrid-daily-1950}\begin{center}\scalebox{0.7}{\begin{tabular}{l*{3}{c}} \hline \hline
				Joint hypothesis tests&Wald test&p-value\\ \hline
				All quantiles (q05, q10,...,q90, q95)&77.046&0.000 \\
				Lower quantiles (q05, q10, q20, q30)           &1.360&0.715 \\
				Medium quantiles (q40, q50, q60)         &2.036&0.361 \\
				Upper quantiles (q70, q80, q90, q95)          &3.944&0.268 \\
				Lower-Medium quantiles (q05, q10, q20, q30, q40, q50, q60)          &6.707&0.349 \\
				Medium-Upper quantiles (q40, q50, q60, q70, q80, q90, q95)          &31.822&0.000 \\
				Lower-Upper quantiles (q05, q10, q20,q30, q70, q80, q90, q95 )          &74.967&0.000 \\
				\hline
				Spacing hypothesis&Trend-coeff.&p-value\\ \hline
				q50-q05                            &0.005&0.505 \\
				q95-q50&0.023&0.000 \\
				q05-q95                            &-0.028&0.000 \\
				q75-q25  (iqr)                          &0.027&0.000 \\
				\hline \hline \end{tabular}}\end{center}
	\begin{tablenotes}
		\textit{Note}:   Annual distributional characteristics (quantiles) of temperature. The top panel shows the Wald test of the null hypothesis of equality
		of trend coefficients for a given set of characteristics. In the bottom panel, the TT is applied to the difference between two
		representative quantiles.
\end{tablenotes}\end{table}

\begin{table}\caption{Co-trending analysis (Madrid-Retiro daily data, AEMET, 1970-2019)}\label{Tab-cotrend-Madrid-daily-1970}\begin{center}\scalebox{0.7}{\begin{tabular}{l*{3}{c}} \hline \hline
				Joint hypothesis tests&Wald test&p-value\\ \hline
				All quantiles (q05, q10,...,q90, q95)&81.371&0.000 \\
				Lower quantiles (q05, q10, q20, q30)           &0.424&0.935 \\
				Medium quantiles (q40, q50, q60)         &8.111&0.017 \\
				Upper quantiles (q70, q80, q90, q95)          &3.214&0.360 \\
				Lower-Medium quantiles (q05, q10, q20, q30, q40, q50, q60)          &45.687&0.000 \\
				Medium-Upper quantiles (q40, q50, q60, q70, q80, q90, q95)          &18.851&0.004 \\
				Lower-Upper quantiles (q05, q10, q20,q30, q70, q80, q90, q95 )          &71.094&0.000 \\
				\hline
				Spacing hypothesis&Trend-coeff.&p-value\\ \hline
				q50-q05                            &0.036&0.004 \\
				q95-q50&0.017&0.051 \\
				q05-q95                            &-0.053&0.000 \\
				q75-q25  (iqr)                          &0.040&0.000 \\
				\hline \hline \end{tabular}}\end{center}
	\begin{tablenotes}
		\textit{Note}:   Annual distributional characteristics (quantiles) of temperature. The top panel shows the Wald test of the null hypothesis of equality
		of trend coefficients for a given set of characteristics. In the bottom panel, the TT is applied to the difference between two
		representative quantiles.
\end{tablenotes}\end{table}

\begin{table}[h!]\caption{Amplification hypothesis (Madrid daily data, AEMET 1950-2019)}\label{Tab-amplif-Madrid-1950}\begin{center}\scalebox{0.8}{\begin{tabular}{l*{5}{c}} \hline \hline
				periods/variables&1950-2019&1970-2019&1950-2019&1970-2019\\ \hline
				& \multicolumn{2}{c}{Inner}& \multicolumn{2}{c}{Outer}\\  \hline
				q05&0.66&0.43&0.83&0.56\\
				& (0.993)& (1.000)& (0.802)& (0.990) \\
				q10&0.58&0.42&0.73&0.54\\
				& (1.000)& (1.000)& (0.974)& (1.000) \\
				q20&0.66&0.53&0.81&0.65\\
				& (1.000)& (1.000)& (0.961)& (0.999) \\
				q30&0.72&0.74&0.94&0.90\\
				& (1.000)& (0.996)& (0.758)& (0.836) \\
				q40&0.90&1.02&1.15&1.21\\
				& (0.887)& (0.436)& (0.072)& (0.041) \\
				q50&1.08&1.29&1.38&1.53\\
				& (0.188)& (0.001)& (0.001)& (0.000) \\
				q60&1.14&1.31&1.44&1.54\\
				& (0.040)& (0.000)& (0.000)& (0.000) \\
				q70&1.22&1.23&1.46&1.38\\
				& (0.012)& (0.019)& (0.000)& (0.002) \\
				q80&1.45&1.36&1.70&1.52\\
				& (0.000)& (0.003)& (0.000)& (0.002) \\
				q90&1.31&1.29&1.48&1.38\\
				& (0.004)& (0.041)& (0.005)& (0.064) \\
				q95&1.31&1.33&1.46&1.39\\
				& (0.001)& (0.021)& (0.007)& (0.073) \\
									\hline \hline \end{tabular}}\end{center}
			\begin{tablenotes}
		\textit{Note}: OLS estimates and HAC p-values of the t-statistic of testing $H_{0}: \beta_{i}=1$ versus $H_{a}: \beta_{i}>1$ in the regression: $C_{it}=\beta _{i0}+\beta _{i1} mean_{t}+\epsilon_{it}$. $mean$ refers to the average of the Madrid or Spanish temperature distribution for the ``inner'' and ``outer''cases, respectively.
\end{tablenotes}\end{table}

\clearpage
\subsection{Barcelona-Fabra}
\begin{figure}[h!]
	\begin{center}
		\caption{Characteristics of temperature data in Barcelona-Fabra (AEMET daily data, 1950-2019)} \label{fig-char-daily-Barcelona-1950}
		\includegraphics[scale=0.6]{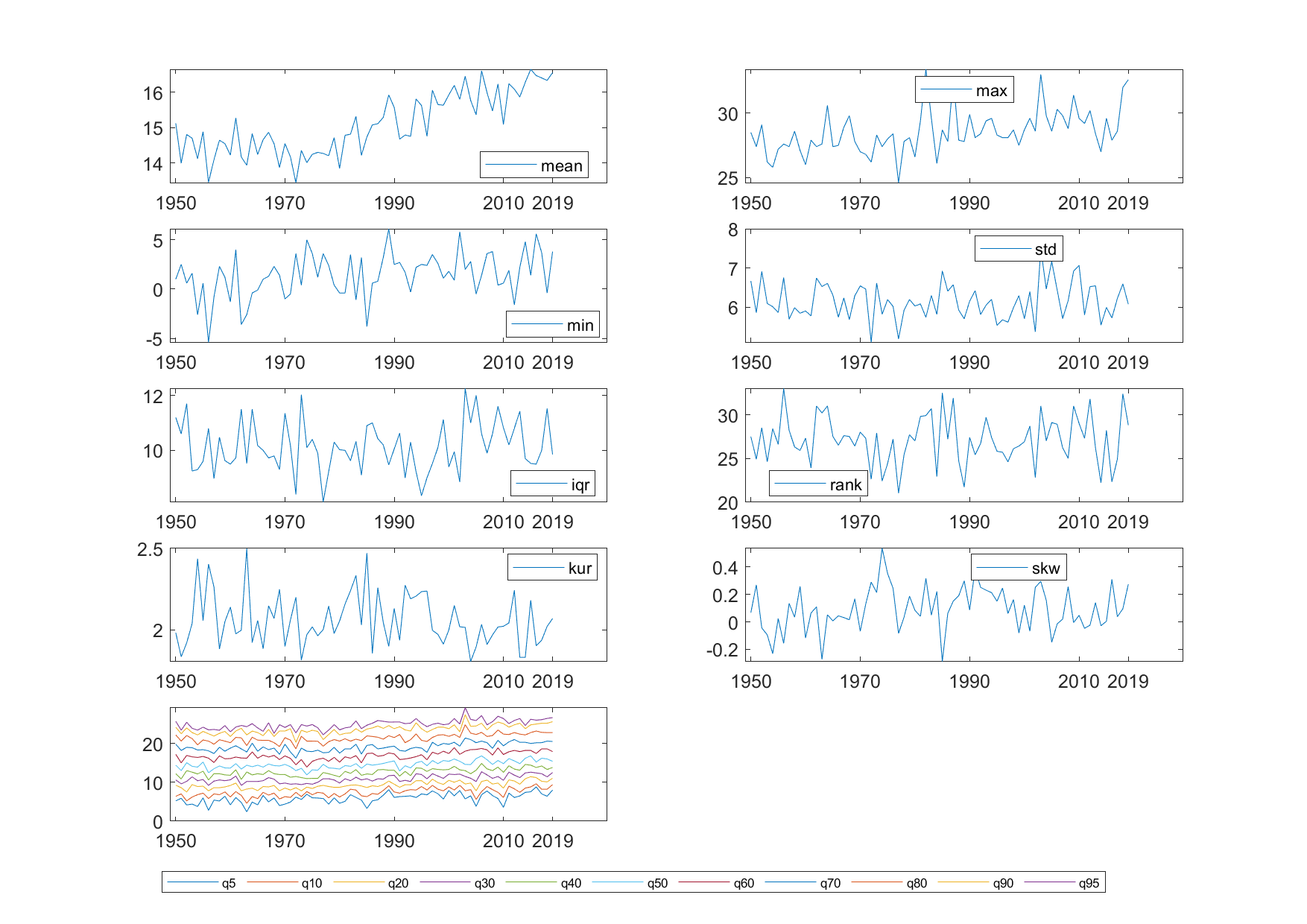}
	\end{center}      	
\end{figure}

\begin{table}\caption{Trend acceleration hypothesis (Barcelona, daily data, AEMET, 1950-2019)}\label{Tab-1950-Barcelona-daily-rec-acc}\begin{center}\scalebox{0.5}{\begin{tabular}{l*{5}{c}} \hline \hline
				& \multicolumn{2}{c}{Trend test by periods}& \multicolumn{1}{c}{Acceleration test}\\
				names/periods&1950-2019& 1970-2019& 1950-2019, 1970-2019\\ \hline
				mean  &0.0340&0.0512&3.2979\\
				& (0.0000)& (0.0000)& (0.0006) \\
				max   &0.0394&0.0531&0.7280\\
				& (0.0000)& (0.0038)& (0.2339) \\
				min   &0.0397&0.0231&-0.7411\\
				& (0.0011)& (0.2654)& (0.7700) \\
				std   &0.0013&0.0057&0.9146\\
				& (0.6185)& (0.1787)& (0.1810) \\
				iqr   &0.0042&0.0113&0.7351\\
				& (0.4418)& (0.1892)& (0.2318) \\
				rank  &-0.0004&0.0300&0.9299\\
				& (0.9806)& (0.3322)& (0.1770) \\
				kur   &-0.0013&-0.0018&-0.2693\\
				& (0.1555)& (0.2075)& (0.6060) \\
				skw   &0.0011&-0.0022&-1.7869\\
				& (0.2678)& (0.1942)& (0.9619) \\
				q5    &0.0374&0.0358&-0.1381\\
				& (0.0000)& (0.0015)& (0.5548) \\
				q10   &0.0350&0.0385&0.4361\\
				& (0.0000)& (0.0000)& (0.3317) \\
				q20   &0.0317&0.0439&1.7009\\
				& (0.0000)& (0.0000)& (0.0456) \\
				q30   &0.0308&0.0488&2.4813\\
				& (0.0000)& (0.0000)& (0.0072) \\
				q40   &0.0324&0.0537&2.9244\\
				& (0.0000)& (0.0000)& (0.0020) \\
				q50   &0.0325&0.0548&2.7535\\
				& (0.0000)& (0.0000)& (0.0034) \\
				q60   &0.0344&0.0636&3.0915\\
				& (0.0000)& (0.0000)& (0.0012) \\
				q70   &0.0330&0.0583&2.9241\\
				& (0.0000)& (0.0000)& (0.0020) \\
				q80   &0.0357&0.0551&2.4081\\
				& (0.0000)& (0.0000)& (0.0087) \\
				q90   &0.0394&0.0567&2.0957\\
				& (0.0000)& (0.0000)& (0.0190) \\
				q95   &0.0390&0.0525&1.3435\\
				& (0.0000)& (0.0000)& (0.0907) \\
				\hline \hline \end{tabular}}\end{center}
	\begin{tablenotes}
		\tiny{		\textit{Note}: OLS estimates and HAC p-values in parenthesis of the $t_{\beta=0}$ test from regression: $C_{t}=\alpha+\beta t+u_{t}$, for two different time periods. For the acceleration hypothesis we run the system: $C_{t}=\alpha_{1} +\beta_{1} t+u_{t}, \text{   } t=1, ...,s ,..., T, C_{t}=\alpha_{2} +\beta_{2} t+u_{t}, \text{   } t=s+1, ..., T, \text{and test the null hypothesis } \beta_{2}=\beta{1} \text{ against the alternative} \beta_{2}>\beta{1}$. We show the value of the t-statistic and its HAC p-value.}
\end{tablenotes}\end{table}

			\begin{table}\caption{Co-trending analysis (Barcelona-Fabra daily data, AEMET, 1950-2019)}\label{Tab-cotrend-Barcelona-daily-1950}\begin{center}\scalebox{0.7}{\begin{tabular}{l*{3}{c}} \hline \hline
							Joint hypothesis tests&Wald test&p-value\\ \hline
							All quantiles (q05, q10,...,q90, q95)&3.368&0.971 \\
							Lower quantiles (q05, q10, q20, q30)           &1.036&0.792 \\
							Medium quantiles (q40, q50, q60)         &0.073&0.964 \\
							Upper quantiles (q70, q80, q90, q95)          &0.784&0.853 \\
							Lower-Medium quantiles (q05, q10, q20, q30, q40, q50, q60)          &1.171&0.978 \\
							Medium-Upper quantiles (q40, q50, q60, q70, q80, q90, q95)          &1.901&0.929 \\
							Lower-Upper quantiles (q05, q10, q20,q30, q70, q80, q90, q95 )          &2.969&0.888 \\
							\hline
							Spacing hypothesis&Trend-coeff.&p-value\\ \hline
							q50-q05                            &-0.005&0.528 \\
							q95-q50&0.006&0.233 \\
							q05-q95                            &-0.002&0.856 \\
							q75-q25  (iqr)                          &0.004&0.442 \\
							\hline \hline \end{tabular}}\end{center}
				\begin{tablenotes}
					\textit{Note}:   Annual distributional characteristics (quantiles) of temperature. The top panel shows the Wald test of the null hypothesis of equality
					of trend coefficients for a given set of characteristics. In the bottom panel, the TT is applied to the difference between two
					representative quantiles.
			\end{tablenotes}\end{table}
			
			\begin{table}\caption{Co-trending analysis (Barcelona-Fabra daily data, AEMET, 1970-2019)}\label{Tab-cotrend-Barcelona-daily-1970}\begin{center}\scalebox{0.7}{\begin{tabular}{l*{3}{c}} \hline \hline
							Joint hypothesis tests&Wald test&p-value\\ \hline
							All quantiles (q05, q10,...,q90, q95)&13.165&0.215 \\
							Lower quantiles (q05, q10, q20, q30)           &1.904&0.593 \\
							Medium quantiles (q40, q50, q60)         &1.267&0.531 \\
							Upper quantiles (q70, q80, q90, q95)          &0.384&0.943 \\
							Lower-Medium quantiles (q05, q10, q20, q30, q40, q50, q60)          &10.103&0.120 \\
							Medium-Upper quantiles (q40, q50, q60, q70, q80, q90, q95)          &1.642&0.949 \\
							Lower-Upper quantiles (q05, q10, q20,q30, q70, q80, q90, q95 )          &9.693&0.207 \\
							\hline
							Spacing hypothesis&Trend-coeff.&p-value\\ \hline
							q50-q05                            &0.019&0.192 \\
							q95-q50&-0.002&0.821 \\
							q05-q95                            &-0.017&0.241 \\
							q75-q25  (iqr)                          &0.011&0.189 \\
							\hline \hline \end{tabular}}\end{center}
				\begin{tablenotes}
					\textit{Note}:   Annual distributional characteristics (quantiles) of temperature. The top panel shows the Wald test of the null hypothesis of equality
					of trend coefficients for a given set of characteristics. In the bottom panel, the TT is applied to the difference between two
					representative quantiles.
			\end{tablenotes}\end{table}
			
\begin{table}[h!]\caption{Amplification hypothesis (Barcelona daily data, AEMET 1950-2019)}\label{Tab-amplif-Barcelona-1950}\begin{center}\scalebox{0.8}{\begin{tabular}{l*{5}{c}} \hline \hline
				periods/variables&1950-2019&1970-2019&1950-2019&1970-2019\\ \hline
				& \multicolumn{2}{c}{Inner}& \multicolumn{2}{c}{Outer}\\
				q05&0.99&0.76&1.19&0.87\\
				& (0.523)& (0.918)& (0.225)& (0.720) \\
				q10&0.90&0.79&1.10&0.94\\
				& (0.824)& (0.980)& (0.263)& (0.668) \\
				q20&0.89&0.85&1.09&1.04\\
				& (0.931)& (0.964)& (0.192)& (0.318) \\
				q30&0.96&0.98&1.22&1.25\\
				& (0.813)& (0.585)& (0.000)& (0.000) \\
				q40&0.99&1.04&1.27&1.33\\
				& (0.570)& (0.300)& (0.000)& (0.000) \\
				q50&1.01&1.07&1.27&1.32\\
				& (0.466)& (0.224)& (0.002)& (0.003) \\
				q60&1.09&1.23&1.29&1.42\\
				& (0.175)& (0.005)& (0.014)& (0.001) \\
				q70&1.09&1.17&1.26&1.31\\
				& (0.128)& (0.012)& (0.022)& (0.008) \\
				q80&1.06&1.04&1.22&1.17\\
				& (0.191)& (0.338)& (0.052)& (0.117) \\
				q90&1.09&1.08&1.22&1.20\\
				& (0.125)& (0.241)& (0.047)& (0.121) \\
				q95&1.06&1.03&1.16&1.12\\
				& (0.304)& (0.432)& (0.192)& (0.298) \\
					\hline \hline \end{tabular}}\end{center}
\begin{tablenotes}
\textit{Note}: OLS estimates and HAC p-values of the t-statistic of testing $H_{0}: \beta_{i}=1$ versus $H_{a}: \beta_{i}>1$ in the regression: $C_{it}=\beta _{i0}+\beta _{i1} mean_{t}+\epsilon_{it}$. $mean$ refers to the average of the Barcelona or Spanish temperature distribution for the ``inner'' and ``outer''cases, respectively.
\end{tablenotes}\end{table}

\end{document}